
\tolerance=10000
\magnification=1200
\raggedbottom

\baselineskip=15pt
\parskip=1\jot

\def\sk{\vskip 3\jot}

\def\heading#1{\vskip3\jot{\noindent\bf #1}}
\def\label#1{{\noindent\it #1}}
\def\QED{\hbox{\rlap{$\sqcap$}$\sqcup$}}


\def\ref#1;#2;#3;#4;#5.{\item{[#1]} #2,#3,{\it #4},#5.}
\def\refinbook#1;#2;#3;#4;#5;#6.{\item{[#1]} #2, #3, #4, {\it #5},#6.} 
\def\refbook#1;#2;#3;#4.{\item{[#1]} #2,{\it #3},#4.}


\def\({\bigl(}
\def\){\bigr)}

\def\Tr{{\rm Tr}} 

\def\Ex{{\rm Ex}} 
\def\Var{{\rm Var}}

\def\ru{${\bf repeat}\, \ldots\, {\bf until}$} 

\def\incyc{{\rm\ in\ }} 
\def\insamecyc{{\rm\ in\ same\ }}
\def\indisjcyc{{\rm\ in\ disjoint\ }}

\def\Ei{{\rm Ei}} 

{ 
\pageno=0
\nopagenumbers

\vskip1.125in
\centerline{\bf Topological Characteristics of Random Surfaces}
\centerline{\bf Generated by Cubic Interactions}
\sk

\centerline{Nicholas Pippenger
\footnote{*}{The work reported here was supported by
an NSERC Research Grant and a Canada Research Chair.}}
\centerline{({\tt nicholas@cs.ubc.ca})}
\centerline{Department of Computer Science}
\sk

\centerline{Kristin Schleich
\footnote{$\dagger$}{The work reported here was supported by
an NSERC Research Grant.}}
\centerline{({\tt schleich@physics.ubc.ca})}
\centerline{Department of Physics and Astronomy}
\sk

\centerline{The University of British Columbia}
\centerline{Vancouver, British Columbia V6T 1Z4}
\centerline{CANADA}
\vskip0.5in

\noindent{\bf Abstract:}
We consider random topologies of surfaces generated by cubic interactions.
Such surfaces arise in various contexts in
$2$-dimensional quantum gravity and as world-sheets
of 
string theory.
Our results are most conveniently expressed in terms of a parameter
$h = n/2 + \chi$, where $n$ is the number of interaction vertices and
$\chi$ is the Euler characteristic of the surface.
Simulations and results for similar models suggest that
$\Ex[h] = \log (3n) + \gamma + O(1/n)$ and
$\Var[h] = \log (3n) + \gamma - \pi^2/6 + O(1/n)$.
We prove rigourously that 
$\Ex[h] = \log n + O(1)$ and
$\Var[h] = O(\log n)$.
We also derive results concerning a number of other characteristics of
the topology of these random surfaces.
\vfill\eject
} 

\heading{1.  Introduction}

Many years ago, Wheeler [1] argued that in a theory of quantum gravity, large fluctuations in 
curvature at small distance scales, mandated by the uncertainty principle, would give rise to
corresponding fluctuations in the topology of spacetime. Thus, though spacetime at large distance
scales appears to have the simple topology of a ball in Euclidean space, at small distance scales 
its topology may be that of a complicated and dynamically changing space-time foam. Some of the 
consequences of space-time foam have been explored by Hawking [2,3] and others
 (see for example Carlip [4,5]) in the context of Euclidean quantum gravity.
Although this work has shed light on the qualitative aspects of topological fluctuations,
it is difficult to obtain quantitative results,
such as the probability distributions of topological invarants,
in four or even three dimensions in other than
very special cases.
Indeed, even recognizing topological equivalence (diffeomorphism) is undecidable for
four-manifolds (see Markov [6], Boone {\it et al.\/} [7] and,
for an overview of this issue and its relationship to quantum gravity,
Schleich and Witt [8]), and an open
problem for three-manifolds. 
On the other hand, two-dimensional topology is sufficiently
tractible that quantitative results can easily be obtained for a variety of models.
Much of this previous work has focused on models in which the 
contributions from the two
dimensional topology enters into the theory weighted in a prescribed fashion by other fields. In contrast,
this paper will explore a two-dimensional model giving rise to a probability distribution on
the topology alone, independent of any geometric or dynamical considerations.

Hermitian matrix models provide a prominent example in which the contribution
of two dimensional topology to quantum amplitudes appears  in conjunction with that of other fields. The general form of the free energy for such a  model is given by
$$\log Z = \log \int \exp -\Tr\(V(\Phi)\) \, d\Phi,$$
where the integral is over all $N\times N$ Hermitian matrices $\Phi$, and $V(\Phi)$ is a 
polynomial in $\Phi$.
The large-$N$ limit, in which $N\to\infty$ with the coefficients of $V$ fixed, has been shown
by 't~Hooft [9, 10] to be dominated by the dynamics of the fields on a surface of genus zero
(see also Brezin {\it et al.\/} [11] and Bessis {\it al.\/} [12]).
The double-scaling limit, in which the coupling constant in $V$ varies in a prescribed way as
$N\to\infty$, includes contributions to the partition function from surfaces of all genera
(see Brezin and  Kazakov [13], Douglas and Shenker [14] and Gross and Migdal [15, 16]).
One such model is equivalent to Euclidean quantum gravity in two dimensions,
which corresponds to a probability distribution in which 
each closed compact connected two-dimensional
manifold is weighted by the exponential of its Euler characteristic
(see also the survey of Di Francesco {\it et al.\/} [17]).

String theory provides another example of a theory in which contribution of the topology of two-dimensional surfaces to quantum amplitudes enters coupled to their geometry, as induced by an embedding in a higher-dimensional space-time.
In perturbative string theory, interacting strings have higher-genus worldsheets
(see Polyakov [18] and Alvarez [19], as well as Polchinski [20] (pp.~86--90) and the references
therein).
In string field theory, Witten [21] proposed a cubic interaction for open strings for which
these worldsheets are two-dimensional surfaces with boundary (see also Horowitz {\it et al.\/}
[22]). 
Note that string theories have natural connections to matrix models;
for example, topological strings provide a natural connection between supersymmetry 
in four dimensions and matrix models (see Ooguri and Vafa [23]).

There are also a number of approaches that treat topology in a discrete way. In particular,
Regge [24] 
introduced  Regge calculus, in which a discrete triangulation forms the
framework underlying a piecewise flat approximation to a continuous geometry. This approach has
proven to be a useful framework for the study of many issues in classical and quantum gravity in
two and more dimensions
(see the surveys of Williams and Tuckey [25] and Williams [26]). A special instance of this
approach, dynamical triangulations, has also been extensively used to investigate properties
of quantum gravity (see also Ambj\o rn {\it et al.\/} [27]). Interestingly enough, in two dimensions,
Weingarten [28], Ambj\o rn {\it et al.\/} [29], Frohlich [30],  David [31, 32]
and Kazakov {\it et al.\/} [33]
have shown that  certain dynamical triangulation models have deep connections with string theory. 
All of these approaches have the property that the topology is prescribed {\it a priori}, so that only quantities related to 
geometry can be  predicted by the theory. However, there are some generalizations of 
Regge calculus that lead to theories that can allow consideration of different topologies. Ponzano and Regge [34] introduced a particular discretization of geometry through spin variables (elements of $SU(2)$) associated with the edges of the triangulation.
Their extension is specific to three dimensions, but work of Penrose [35, 36, 37] and others
(see Barrett and Crane [38] and Baez [39, 40]) on spin networks has
generalized it to higher dimensions by associating spin variables (in $SU(2)$ and other groups)
with two-dimensional simplices. Certain formulations of spin networks allow consideration of the contribution of different topologies. However, as in string theory and matrix models, topological
 and spin variable contributions enter together  in the computation of physical quantities.

In an interesting approach closely related to our model, Hartle [41] has formulated a two dimensional Regge calculus model in which one sums over topologies as well as geometries in computing amplitudes, and which treats the topology in a discrete fashion in a form separable from its geometry. Hartle then considers the contribution of pseudomanifolds to quantum amplitudes by calculating the probability distribution of pseudomanifolds for small triangulations.

The model explored in this paper shares with string theory a cubic interaction,
corresponding to graphs in which each vertex is incident with three edges. Our model is also related to  a matrix model with $N=1$ and $V$ a cubic polynomial. However, unlike these
examples, our model focusses on the discrete space of topologies of surfaces, ignoring
geometric and other structure. As elucidated in Section 2, our model is entirely discrete, and can be formulated combinatorially in terms of graphs or permutations. Our model shares with Hartle's model this discrete nature. However, it differs from Hartle's model in three aspects:  1) Hartle's model  assigns positive probabilities to pseudomanifolds whereas we consider only manifolds, 2) Hartle's model assigns equal probabilities to all pseudomanifolds with a given number of vertices whereas we assign equal probabilites to all manifolds with a given number of triangles, and
3) Hartle's model requires that the pseudomanifolds be simplicial whereas we do not require this of our triangulations. These differences allow us to examine the probability distribution of our model for  large triangulations.


\heading{2.  Probabilistic Models}

We shall focus our attention on one particular probability distribution, but there are several
mathematical models giving rise to this distribution. 
The simplest of these models, the {\it quotient model}, produces a random orientable
surface as the quotient surface of a number of solid triangles.
We shall refer to the elements of the boundaries of these triangles as {\it corners\/} and {\it arcs},
rather than as vertices and edges (since we shall reserve the terms vertices and edges for elements
in the thin graph model presented below).
Take the $3n$ arcs of an even number $n$ of 
oriented triangles and identify them in pairs,
respecting the orientation, so that the resulting surface is orientable, and
with all $(3n-1)!! = (3n-1)\cdot (3n-3) \cdots 3\cdot 1$ pairings being equally likely.
(These pairing may identify two arcs from the same triangle, or more than one pair of arcs between 
two triangles.)
The surface is triangulated with $n$ triangles, and their $3n$ arcs before identification
become $3n/2$ arcs after identification.
Let the random variable $h$ denote the number of corners after identification, and let
$\chi$ denote the Euler characteristic of the surface.
Then we have $\chi = h - 3n/2 + n$, so that $h = \chi + n/2$.
The distribution of $h$ will be the main object of study in this paper,
but studying $h$ is equivalent to studying $\chi$.
We note that, since $\chi$ is even, $h$ has the same parity as $n/2$.
We shall see in Section 3 that with probability 
$1 - 5/18n + O(1/n^2)$ the surface is connected, and thus consists of a single component.
In this case, the genus $g$ of the surface is given by
$g = 1 - \chi/2 = 1 + n/4 - h/2$.
Thus in this case studying $h$ is equivalent to studying $g$.

A variant of the quotient model is the {\it fat-graph model}.
Take $n$ triangles as before, but instead of identifying their arcs in pairs,
join them in pairs with rectangular {\it ribbons}, again in such a way that the resulting surface
is orientable, and again with all $(3n-1)!!$ pairings being equally likely.
The resulting surface in this case will have a non-empty boundary,
comprising one or more {\it boundary cycles}.
If we shrink each boundary cycle to a point (or if we cap off each boundary cycle with an
appropriate polygon), we obtain a surface topologically equivalent to that produced by the
quotient model for the same pairing.
The points produced by shrinking boundary components correspond to the corners after
identification in the quotient model, so that the number of boundary components is $h$.
We shall exploit this correspondence in Sections 4 and 5 to study $h$ by analyzing an algorithm
that traces out the boundary cycles in the fat-graph model.

Another variant of the quotient model is the {\it thin-graph model}.
In the quotient surface, install a {\it vertex\/} corresponding to  each triangle 
and install an {\it edge\/} joining two vertices for each pair of arcs of the  
corresponding triangles that are identified.
(Alternatively, one may shrink each triangle in the fat graph to a vertex, and shrink
each ribbon to an edge.
In any case, the graph may have ``loops'' (edges joining vertices to themselves) and
``slings'' (multiple edges joining a single pair of vertices).)
The result is a cubic graph (each vertex has degree $3$), with an additional structure:
the three edges incident with each vertex have one of two possible cyclic orderings.
As was observed by Heffter [42],
these cyclic orderings allow the quotient surface or fat graph to be reconstructed
from the thin graph.
The thin graph may be regarded as a bubble diagram for a cubic interaction,
and it is this viewpoint that forms the physical basis for our models.
The thin graph, even without the cyclic ordering structure, reveals the number of connected
components of the random surfaces in the quotient and fat-graph models.
We shall exploit this fact in Section 3 to estimate the probability that
these surfaces are connected.

A final variant of these models is the {\it permutation model}.
Here we ignore surfaces and graphs and simply consider a probability distribution
on permutations.
Let $\varrho$ be a permutation of $3n$ elements having cycle structure $[2^{3n/2}]$
(say $(12)\,(34)\cdots (3n-1\;3n)$), and let $\sigma$ be a permutation having cycle structure
$[3^n]$ (say $(123)\,(456)\cdots(3n-2 \; 3n-1 \; 3n)$).
Let $\pi$ be a random permutation, with all $(3n)!$ permutations being equally likely.
Then $\pi\,\varrho\,\pi^{-1}$ is a random permutation uniformly distributed over the permutations
with cycle structure $[2^{3n/2}]$.
If we think of the cycles of $\sigma$ as corresponding to the triangles of the fat graph
model and those of $\pi\,\varrho\,\pi^{-1}$ as corresponding to the ribbons, we see that
the cycles of $\pi\,\varrho\,\pi^{-1}\,\sigma$ correspond to the boundary cycles in the fat graph.
We shall exploit this correspondence in Section 6 to to study the probability that $h=1$.

For $n=2$ triangles, there are $5!! = 15$ possible pairings, but they fall into just
three combinatorial equivalence classes.
(Two pairings are combinatorially equivalent if they lie in the same orbit 
under the action of the symmetry group
that permutes the triangles and cyclically permutes their boundary elements (without changing their
orientation).)
Three representative pairings for these equivalence classes are shown in Figure 1; alongside them are shown
their corresponding thin graphs (or their fat graphs with very small triangles and very long and slender
ribbons). 
The arrows indicate the cyclic orderings of the edges at each vertex.
These three representatives are in classes of sizes $9$, $3$ and $3$ (from top to bottom);
the upper two yield spheres ($h=3$), while the lowest one yields a torus ($h=1$).
Thus the probability of a sphere is $(9+3)/15 = 4/5$, while the probability of a torus is $3/15 = 1/5$.

There are two other mathematical models that give rise to different probability distributions,
but which yield results similar to those we have observed for
random surfaces generated by cubic interactions.
The first of these is simply a random permutation of $n$ elements, with all
$n!$ permutations being equally likely.
Let $h'$ denote the number of cycles in such a random permutation.
We note that $h'$ may take any value from $1$ to $n$.
The probability distribution of $h'$ is given by the 
generating function ${\xi + n - 1\choose n}$, in which $\Pr[h'=k]$ is the coefficient of 
$\xi^k$.
We have $\left({d\over d\xi}{\xi + n - 1\choose n}\right)_{\xi=1} = H_n$,
where $H_n = \sum_{1\le k\le n}{1\over k} = \log n + \gamma + O(1/n)$ and
$\gamma = 0.5772\ldots$ is Euler's constant, and
$\left({d^2\over d\xi^2}{\xi + n - 1\choose n}\right)_{\xi=1} = H_n^2 - H_n^{(2)}$,
$H_n^{(2)} = \sum_{1\le k\le n}{1\over k^2} = \pi^2/6 + O(1/n)$ and
$\pi = 3.14159\ldots$ is the circular ratio.
These results yield
$\Ex[h'] = \log n + \gamma + O(1/n)$
and
$\Var[h'] = \log n + \gamma  - \pi^2/6 + O(1/n)$.
We also note that $\Pr[h'=1] = 1/n$.

Another analogous model was introduced by Harer and Zagier [43].
Instead of starting with $n$ triangles, each having $3$ arcs,
they start with a single polygon having $2n$ arcs, and again identify
the arcs in pairs to obtain an orientable surface, with all 
$(2n-1)!!$ pairings being equally likely.
Let $h''$ denote the number of equivalence classes of corners in the 
resulting quotient surface 
(or equivalently the number of boundary cycles in the
resulting fat graph).
We note that $h''$ has the opposite parity from $n$.
They showed that the generating function for $h''$ is
${1\over 2}\sum_{l+m=n+1} {\xi\choose l}{\xi+m-1\choose m}$
by expressing this generating function as a $\xi$-fold integral over
a Gaussian Hermitian ensemble.
(Subseqently, Penner [44] derived this result using perturbative series,
and Itzykson and Zuber [45] derived in two further ways, one using group representations
and another expoiting an analogy with the second quantization of the harmonic oscillator.)
We have $\left({d\over d\xi} {\xi\choose l}\right)_{\xi=1} = 1$ if $l=1$
and $(-1)^l/l(l-1)$ otherwise, and
$\left({d^2\over d\xi^2} {\xi\choose l}\right)_{\xi=1} = 0$ if $l=1$
and $2(-1)^{l-1}(H_{l-2}-1)/l(l-1)$ otherwise.
These results yield
$\Ex[h''] = \log (2n) + \gamma + O(1/n)$
and
$\Var[h''] = \log (2n) + \gamma  - \pi^2/6 + O\((\log n)/n\)$.
We also note that $\Pr[h''=1] = 1/(n+1)$. 

We have conducted an empirical study of $10{,}000$ random surfaces, each constructed from
$80{,}000$ triangles.
The sample mean of $h$ was $13.1092$ and the sample variance was $11.3216\ldots\,$.
Since $\log(240{,}000) + \gamma = 12.9656\ldots$ and 
$\log(240{,}000) + \gamma - \pi^2/6 = 11.3206\ldots\,$,
these results strongly suggest the conjecture that
$$\Ex[h] = \log(3n) + \gamma + O\left({1\over n}\right)$$
and
$$\Var[h] = \log(3n) + \gamma - {\pi^2\over 6}+ O\left({1\over n}\right).$$
These conjectures may be compared with the results 
for the Harer-Zagier model:
in each case the argument of the logarithm is 
the total number of arcs of the original polygons,
and all remaining constant terms are identical.
We have not been able to verify these conjectures at the stated level of precision,
but in Section 4 we shall prove that
$$\Ex[h] = \log n + O(1)$$\
and in Section 5 we shall prove that
$$\Var[h] = O(\log n).$$
Since $\Var[h]/\Ex[h]^2 \to 0$ as $n\to\infty$,
this implies that the distribution of $h$ is strongly concentrated about its mean.

In Section 6 we shall study the probability that $h=1$.
We use a technique based on representations of the symmetric group to show that
show that
$$\Pr[h=1] = \cases{
0, &if $n/2$ even; \cr
& \cr
{\displaystyle 2\over \displaystyle 3n} + 
O\left({\displaystyle 1\over \displaystyle n^2}\right), &if $n/2$ odd. \cr
}$$
This result may be compared with that for the Harer-Zagier model:
in each case the probability, when it does not vanish, is asymptotic to the reciprocal
of the number of edges in the thin graph.

In Section 7 we shall present a classification of the boundary cycles in the fat-graph
model according to their ``self-interactions'',
and give a heuristic estimation of the expected number of ``simple'' cycles of 
various orders.
In Sections 8 and 9 we give exact expressions, and rigourously derive asymptotic estimates,
for the number of simple cycles having the 
two lowest orders in this classification.

In the remainder of this paper, the variable $n$ will always denote the number of triangles, and the random
variable $h$ will always denote the number of boundary cycles in the fat graph; all other variables may be
used with different meanings from section to section.
Unless otherwise indicated, the notation $O(\cdots)$ and $\Omega(\cdots)$ will refer to asymptotic behaviour
as $n$ tends to infinity through even integers.
\sk

\heading{3.  The Probability of Connectedness} 

\label{Theorem 3.1:}
Let $c$ be the number of connected components of a random 
surface obtained from the quotient model with $n$ triangles. 
Then
$$\Pr[c=1] = 1 - {5\over 18n} + O\left({1\over n^2}\right).$$ 

\label{Proof:}
We shall show that
$$\Pr[c\ge 2] = {5\over 18n} + O\left({1\over n^2}\right), \eqno(3.1)$$ 
which is equivalent to the theorem.
We shall work with the thin graph, which has the same number of connected
components as the quotient surface.

First we show that
$$\Pr[c\ge 2] \le {5\over 18n} + O\left({1\over n^2}\right). \eqno(3.2)$$ 
We may assume $n\ge 4$, since since all cubic graphs on $2$ vertices are
connected. 
Let $V$ be the  set of $n$ vertices of the thin graph.
Say that a subset $P$ of the vertices of the thin graph is 
{\it closed\/} of no edge of the thin graph joins a vertex in $P$ to
one in $V\setminus P$.
(Thus a closed set is one that comprises one or more connected components
of the thin graph.)
The event $c\ge 2$ is equivalent to the existence of a closed set $P$
with $\emptyset\not=P\not=V$.
If $P$ is such a closed set, then so is $V\setminus P$, and at least one
of these sets must contain at most $n/2$ vertices.
Thus
$$\Pr[c\ge 2] \le \sum_{\#P \le n/2} \Pr[P{\rm \ closed}].$$
A closed set must contain an even number of vertices,
and there are ${n\choose 2k}$ sets containing $2k$ vertices.
The probability that a given set $P$ containing $2k$ vertices is closed is
$${(6k-1)!! \, (3n-6k-1)!! \over (3n-1)!!},$$ 
since there are $(6k-1)!!$ ways 
of constructing a thin graph on $P$ and $(3n-6k-1)!!$ ways of constructing
a thin graph on $V\setminus P$.
Thus we have
$$\Pr[c\ge 2] \le \sum_{1\le k\le n/4} F_k,$$
where
$$F_k = 
{(6k-1)!! \, (3n-6k-1)!! \over (3n-1)!!}{n\choose 2k}.$$
We shall show that $F_k$ is a non-increasing function of $k$ for
$1\le k\le n/2$.
The ratio of consecutive terms,
$$R_k = {F_{k+1}\over F_k} = 
{(6k+5)(6k+1)(n-2k)\over (3n-2k-1)(3n-2k-5)(2k+2)},$$
is unity when the difference between its numerator and denominator,
$$S_k = (6k+5)(6k+1)(n-2k) - (3n-2k-1)(3n-2k-5)(2k+2),$$
vanishes.
This cubic polynomial in $k$ has roots at $k_0 = n/4 - 1/2$ and
$$k_\pm = {3n - 6 \pm \sqrt{9n^2 + 36n + 16}\over 12}.$$
Since $S_k \sim -144k^3$ for $k\to\pm\infty$, $S_k \le 0$
and thus $R_k \le 1$ for $k_- \le k \le k_0$.
Since $n\ge 4$, we have $144n \ge 308$, which implies that $k_- \le 1$.
Thus we have $F_{k+1} \le F_k$ for $1\le k$ and $k+1\le n/4$.
This implies
$$\Pr[c\ge 2] \le F_1 + F_2 + (n/2 - 2)F_3.$$
Since $F_1 = 5/18n + O(1/n^2)$, $F_2 = O(1/n^2)$ and $F_3 = O(1/n^3)$,
this inequality yields (3.2).

Finally we show that
$$\Pr[c\ge 2] \ge {5\over 18n} + O\left({1\over n^2}\right). \eqno(3.3)$$
To do this we focus our attention on closed sets containing $2$ vertices.
We have
$$\Pr[c\ge 2] \ge \sum_P \Pr[P{\rm \ closed}] - 
\sum_{P,Q} \Pr[P{\rm \ closed}, Q{\rm \ closed}],$$
where the first sum is over all $P$ with $\#P = 2$, and the second sum
is over all unordered pairs of distinct sets $P$ and $Q$ with
$\#P = \#Q = 2$.
The first sum contains ${n\choose 2} = n^2/2 + O(n)$ terms, each equal
to $5!! \, (3n-7)!! / (3n-1)!! = 5/9n^3 + O(1/n^4)$.
For the second term, $\Pr[P{\rm \ closed}, Q{\rm \ closed}]$ vanishes
unless $P$ and $Q$ are disjoint.
Thus the second sum contains ${n\choose 2}{n-2\choose 2}/2 = O(n^4)$
non-vanishing
terms, each equal to $5!! \, 5!! \, (3n-13)!! / (3n-1)!! = O(1/n^6)$.
Substituting these results in (4) yields (3).

Inequalities (3.2) and (3.3) yield (3.1), which completes the proof of the
theorem.
\QED
\sk

\heading{4.  The Expected Number of Cycles}

Our main result is an estimate for the parameter $h$, regarded as the number of boundary cycles in
the fat-graph model with $n$ triangles.

\label{Theorem 4.1:} 
As $n$ tends to infinity through even integers, we have
$$\Ex[h] = \log n + O(1).$$

Let $c$ be one of the $3n$ corners of the original $n$ triangles.
Let $p_k$ denote the probability that $c$ lies in a boundary cycle of length $k$.

\label{Proposition 4.2:}
$$\Ex[h] = 3n \sum_{1\le k\le 3n} {p_k \over k}.$$

\label{Proof:}
Let $h_k$ denote the number of boundary cycles of length $k$.
Then 
$$\Ex[h] = \sum_{1\le k\le 3n} \Ex[h_k].$$
Let $c_k$ denote the number of corners in boundary cycles of length $k$.
Then $c_k = k\, h_k$, so that
$$\Ex[h] = \sum_{1\le k\le 3n} {\Ex[c_k] \over k}. \eqno(4.1)$$
The probability distribution is invariant under a group of symmetries that includes
permutations of the triangles and cyclic permutations of the corners of each triangle.
Since this group acts transitively on the $3n$ corners, we have
$\Ex[c_k] = 3n\,p_k$.
Substituting this into (4.1) yields the proposition.
\QED

Our next step is to obtain estimates for $p_k$.
This will be done through the following two lemmas.

\label{Lemma 4.3:}
For $1\le k\le n$, we have
$$p_k \le {1\over 3n-2k+1} \left(1 + {k\over 3n-2k+5}\right).$$

\label{Lemma 4.4:}
For $1\le k\le n/2$, we have
$$p_k \ge {1\over 3n} \left(1 - {4k\over 3n-2k+1}\right).$$

We shall prove these two lemmas below.
For now, let us see how they combine to proove the following proposition.

\label{Proposition 4.5:}
For $1\le k\le n/2$, we have
$${3n\,p_k \over k} = {1\over k} + O\left({1\over n}\right)$$
as $n$ tends to infinity through even integers.

\label{Proof:}
From Lemma 4.3, we have
$$\eqalign{
{3n\,p_k\over k}
&\le {3n\over k}\cdot{1\over 3n-2k+1} \left(1 + {k\over 3n-2k+5}\right) \cr
&\le \left({1\over k} + {2\over 3n-2k+1}\right)\left(1 + {k\over 3n-2k+5}\right) \cr
&= \left({1\over k} + O\left({1\over n}\right)\right)\left(1 + O\left({k\over n}\right)\right) \cr
&= {1\over k} + O\left({1\over n}\right). \cr
}$$
From Lemma 4.4, we have
$$\eqalign{
{3n\,p_k\over k}
&\ge {1\over k}\left(1 - {4k\over 3n-2k+1}\right) \cr
&= {1\over k} + O\left({1\over n}\right). \cr
}$$
Combining these bounds yields the proposition.
\QED

This proposition, together with Proposition 4.2, allows us to prove Theorem 4.1.

\label{Proof of Theorem 4.1:}
Since the sum of the lengths of all boundary cycles is $3n$, there can be at most $5$ boundary cycles of
length exceeding $n/2$.
Thus Proposition 4.1 yields
$$\Ex[h] = 3n \sum_{1\le k\le n/2} {p_k \over k} + O(1).$$
Evaluating the sum using Proposition 4.5 yields
$$\eqalign{
\Ex[h]
&= \sum_{1\le k\le n/2} \left({1\over k} + O\left({1\over n}\right)\right) + O(1) \cr
&= \log n + O(1), \cr
}$$
since $\sum_{1\le k\le n/2} {1\over k} = \log n + O(1)$.
\QED

It remains to prove Lemmas 4.3 and 4.4.
To do this, we shall analyze the following randomized
procedure $\it cycle$, which constructs the boundary cycle
containing the corner $c$, and returns its length as $\it cycle(c)$.

Since we are dealing with orientable surfaces, we may regard them as two-sided,
and may imagine one side coloured green and the other coloured orange.
We shall regard the arcs of the triangles as being directed clockwise as seen from the green side.
For any corner $d$ of a triangle $T$, let $d^-$ denote the preceding corner
and $d^+$ the following corner in the boundary of $T$, so that $(d, d^+)$, $(d^+, d^-)$ and 
$(d^-, d)$ are the directed arcs in the boundary of $T$.

Each ribbon installed by the following procedure is a quadrilateral containing four corners
and four arcs.
Two of these arcs come from the triangles joined by the ribbon; the other two will be called {\it links}.
Links will be directed so that the arcs of a ribbon are directed counterclockwise as seen from the green side.
Links are initially {\it unmarked}, but may subsequently be {\it marked\/} to indicate that they are part of
the boundary cycle containing $c$.

The procedure uses a data structure called the {\it urn}.
The urn contains at any time a set of corners.
This set is initialized by putting the $3n$ corners into it.
A specific corner may be {\it removed\/} from the urn, or a random corner may be {\it drawn\/} from the urn;
the corner removed in this way is equally likely to be any of the corners currently in the urn.
\sk

{\obeylines \baselineskip=10pt
{\bf integer procedure} $\it cycle({\bf corner\ } c)$; 
{\bf begin}
\quad{\bf corner} $\it head$, $\it tail$, $\it next$, $\it point$;
\quad{\bf integer} $\it length$;
\quad $\it head:= c$;
\quad $\it tail:= c$;
\quad $\it length:=0$;
\quad put the $3n$ corners into an urn;
\quad{\bf repeat}
\qquad remove $\it head$ from the urn;
\qquad $\it next:=$ draw from the urn;
\qquad install a ribbon with arcs $\it (head, next^-)$, $\it (next^-, next)$, $\it (next, head^-)$
\qquad\quad and $\it (head^-, head)$ in counterclockwise order as seen from the green side,
\qquad\quad and introduce $\it (head, next^-)$ and $\it (next, head^-)$ as unmarked links;
\qquad{\bf while} there is an unmarked link $\it (head, point)$ {\bf do}
\qquad\quad{\bf begin}
\qquad\qquad mark the link $\it (head, point)$;
\qquad\qquad $\it length:=length+1$;
\qquad\qquad $head:=point$
\qquad\quad{\bf end};
\qquad{\bf while} there is an unmarked link $\it (point, tail)$ {\bf do}
\qquad\quad{\bf begin}
\qquad\qquad mark the link $\it (point, tail)$;
\qquad\qquad $\it length:=length+1$;
\qquad\qquad $tail:=point$
\qquad\quad{\bf end};
\quad{\bf until} $\it head = tail$;
\quad{\bf return} $\it length$
{\bf end} \par}
\medskip

The \ru\ statement is analogous to a ${\bf while}\, \ldots\, {\bf do}$ statement,
except that the body of the statement is executed before, rather than after, the condition is tested.

Figure 2 shows the three possibilites for the installation of the first ribbon;
unmarked links are shown dashed, and marked links are shown bold.
The most common case, in which the corner $d$ drawn from the urn
lies on a different triangle from $c$, is shown on the left;
the special cases in which $d=c^-$ and $d=c^+$ are shown on the upper right and lower right, respectively.

The probability $p_k$ that the corner $c$ is in a boundary cycle of length $k$ is simply the
probability that the procedure invocation $\it cycle(c)$ returns the value $k$.
Our estimates for $p_k$ will therefore be based on an analysis of this procedure.

Let us consider the subgraph formed by the corners and the
links at some point in the execution of the procedure.
By the {\it in-degree\/} of a corner we shall mean the number (zero or one) of links directed into it,
and by the {\it out-degree}, the number (also zero or one) directed out of it.
This subgraph then comprises one or more components, each of which is either a {\it path\/}
(a part of a boundary cycle)
or a complete boundary cycle.
Initially, there are $3n$ paths, each of length zero.
As links are introduced, paths may be extended,
merged or closed into cycles.
In a path, the corner with out-degree zero is called the {\it front\/} and the corner with in-degree
zero is called the {\it rear}.
At any time, there is one marked path.
Its front is called the {\it head\/} and its rear is called the {\it tail}.
The variable {\it length\/} keeps track of the length of the marked path.
Initially the head and tail coincide at $c$.
When they again coincide after one or more ribbons have been installed, 
the marked path closes into a cycle,
the procedure terminates by returning the length of this cycle.
We observe that the length of the marked path increases by at least one at each execution fo the \ru\
statement, but may increase by more if previously unmarked paths are merged at its head, or tail, or both.

To analyze the behaviour of the procedure, 
it will be convenient to represent its possible
executions as a tree.
This tree will comprise a number of {\it nodes}, each representing a state of 
execution of the procedure,
joined by {\it branches}, each representing an execution of the body of the 
\ru\ statement.
One node, called the {\it root\/} of the tree, 
corresponds to the initial state of the procedure just before
the first execution of the body of the \ru\ statement.
Other nodes, called
the {\it  leaves\/} of the tree correspond to the final states of the procedure after the last executions
of the body of the \ru\ statement.
Each node other than the root has one or more {\it children}, corresponding to the states reach after
vaious corners are drawn from the urn.
Each node that is not a leaf has a unique {\it parent}, corresponding to the immediately
preceding state in the execution.
The {\it ancestors\/} and {\it descendents\/} of a node are defined in the obvious way.

With each node $K$ we may associate a value $\it depth(K)$ 
(corresponding to the number of installed
ribbons), as well as values $\it head(K)$, $\it tail(K)$ and $\it length(K)$, in the obvious way.
Since each installation of a ribbon adds either one or two marked links, we have
$${\it depth}(K) \le {\it length}(K) \le 2{\it depth}(K).$$
All nodes at the same depth $d$ correspond to states in which the same number $3n-2d$ of corners 
(including the current {\it head\/}) remain in the urn, and thus all of these nodes have the same number 
$3n-2d-1$ of children.
If control reaches some node of the tree, it is equally likely to proceed to each of its children.
We shall say that a node $K$ is {\it shallow\/} if ${\it depth}(K) \le k-1$.
If $K$ is shallow, the probability of proceeding to any particular child of $K$ is
at least $1/3n$ and at most $1/(3n-2k+1)$.
Finally, we shall call an internal node $K$ (that is, a node other than a leaf) {\it double\/} if 
$\it head(K)^- = tail(K)$, and call it {\it single\/} otherwise.
If a node is double, then in proceeding to one of its children both newly added links will be marked,
otherwise only one will be marked.
(In either case, previously added links may also be marked.)

We are now erady to prove Lemmas 4.3 and 4.4.

\label{Proof of Lemma 4.3:}
Let $E_k$ denote the event that execution terminates at a leaf $L$ with $\it length(L) = k$,
so that $p_k = \Pr[E_k]$.
If $K$ is a node other than the root, let $K^*$ denote its parent.
We shall write
$$p_k = \Pr[E_k] = \Pr[E^1_k] + \Pr[E^2_k], \eqno(4.2)$$
where $E^1_k$ (respectively, $E^2_k$) denotes the event that execution terminates at a leaf $L$
such that $\it length(L) = k$ and $L^*$ is single (respectively, double).

Let us first consider an upper bound for $\Pr[E^1_k]$.
If $E^1_k$ occurs at $L$, we shall call $L^*$ a {\it precursor\/} for $E^1_k$.
If the node $K$ is a precursor for $E^1_k$, then (1) $K$ has exactly one child at which $E^1_k$
occurs (this child corresponds to drawing the corner $\it tail(K)^+$ from the urn as $\it next$),
(2) ${\it length}(K) = k-1$ (since proceeding to the child at which $E^1_k$ occurs will add one to 
$\it length$, resulting in $\it length = k$), and (3) $K$ is shallow (since installing a ribbon
increases $\it length$ by at least one, so that ${\it depth}(K) \le {\it length}(K) = k-1$).
Letting $K$ also denote the event that control reaches the node $K$, we have
$$\Pr[E^1_k \mid K] = {1\over 3n - 2{\it depth}(K) - 1} \le {1\over 3n-2k+1}.$$
Along any path from the root to a leaf in the tree, at most one node can be a precursor to 
$E^1_k$ (since $\it length$ strictly increases along any such path).
Thus
$$\sum_{\textstyle{\rm precursor\ } K\atop \textstyle{\rm to\ } E^1_k} \Pr[K] \le 1.$$
The last two bounds together yield
$$\eqalignno{
\Pr[E^1_k]
&= \sum_{\textstyle{\rm precursor\ } K\atop \textstyle{\rm to\ } E^1_k} \Pr[E^1_k \mid K] \, \Pr[K] \cr
&\le {1\over 3n-2k+1} \sum_{\textstyle{\rm precursor\ } K\atop \textstyle{\rm to\ } E^1_k} \Pr[K] \cr
&\le {1\over 3n-2k+1}. &(4.3)\cr
}$$

Next let us consider an upper bound for $\Pr[E^2_k]$.
At a given node $K$, we shall say that a corner $\alpha$ {\it leads to\/} a corner $\beta$ {\it in $l$ steps\/}
if $\alpha$ is the rear and $\beta$ is the front of an unmarked path of length $l$ at $K$.
A corner $\gamma$ such that $\gamma^-$ leads to $\gamma$ in $l$ steps will be called a {\it reflector\/} of size
$l$. 
If $E^2_k$ occurs at a leaf $L$, so that $L^*$ is double, we shall call $L^*$ a {\it precursor\/}
for $E^2_k$.
If $K$ is a precursor for $E^2_k$, then (1) there exists an $l\ge 1$ such that all children of $K$ at
which $E^2_k$ occurs correspond to drawing reflectors of size $l$ from the urn, and
(2) ${\it length}(K) = k-l-2$ (since proceeding to a child at which $E^2_k$ occurs will add $l+2$ to $\it
length$, resulting in $\it length = k$), and (3) ${\it depth}(K) \le k-3$ (since
${\it depth}(K) \le {\it length}(K) \le k-3$).

If $\alpha$ is a reflector of size $l$ at node $K$, and if ${\it length}(K) = k-l-2$, then $K$ will be
called a {\it precursor\/} for $\alpha$.
Let $S^\alpha_K$ be the event that corner $\alpha$ is drawn at node $K$.
Then
$$\Pr[S^\alpha_K \mid K] = {1\over 3n-2{\it depth}(K)-1} \le {1\over 3n-2k+5}. \eqno(4.4)$$

Let $R^\alpha_J$ denote the event that corner $\alpha$ becomes a reflector at some child of node $J$.
For any node $J$, there is at most one corner $\alpha$ 
(namely $\it head(J)^+$) that can become a reflector at a child of $J$,
and if there is such an $\alpha$, then
there is just one child of $J$ (namely the one corresponding to drawing the corner $\beta$ from the urn,
where $\beta$ is the rear of the unmarked path with $\alpha$ as its front) at which $\alpha$ becomes a reflector.
Thus, for any node $J$ of depth at most $k-1$,
$$\sum_\alpha \Pr[R^\alpha_J \mid J] \le {1\over 3n-2{\it depth}(J)-1} \le {1\over 3n-2k+1}. \eqno(4.5)$$

For $E^2_k$ to occur, some corner $\alpha$ must become a reflector of size $l$ at a child of some 
shallow node $J$, and then at some descendant $K$ of $J$ that is a precursor for $\alpha$,
$\alpha$ must be drawn from the urn.
Thus
$$\eqalign{
\Pr[E^2_k] &= 
\sum_{{\textstyle J{\rm \ shallow}}} \Pr[J] \, \sum_\alpha \Pr[R^\alpha_J \mid J] 
\quad \sum_{{\textstyle K {\rm \ precursor\ for\ }\alpha \atop\textstyle K {\rm \ descendant\ of\ }J}}
\Pr[K\mid J] \, \Pr[S^\alpha_K \mid K]. \cr}$$
Using $(4.4)$, we have
$$\eqalign{
\Pr[E^2_k] \le {1\over 3n-2k+5}
\sum_{{\textstyle J{\rm \ shallow}}} \Pr[J] \, &\sum_\alpha \Pr[R^\alpha_J \mid J] 
\quad \sum_{{\textstyle K {\rm \ precursor\ for\ }\alpha \atop\textstyle K {\rm \ descendant\ of\ }J}}
\Pr[K\mid J]. \cr}$$
If $\alpha$ is a reflector of size $l$, a node $K$ can be a precursor for $\alpha$ only if 
${\it length}(K) = k-l-2$.
Since $\it length$ strictly increases along any path descending from $J$, at most one
node on any such path can be a precursor for $\alpha$.
Thus we have
$$\sum_{{\textstyle K {\rm \ precursor\ for\ }\alpha \atop\textstyle K {\rm \ descendant\ of\ }J}}
\Pr[K\mid J] \le 1.$$
This yields
$$\Pr[E^2_k] \le {1\over 3n-2k+5}
\sum_{{\textstyle J{\rm \ shallow}}} \Pr[J] \, \sum_\alpha \Pr[R^\alpha_J \mid J].$$
Using $(4.5)$, we have
$$\Pr[E^2_k] \le {1\over 3n-2k+1}\,{1\over 3n-2k+5}
\sum_{{\textstyle J{\rm \ shallow}}} \Pr[J] .$$
Finally, since there are at most $k$ nodes of depth at most $k-1$ on any path from the root to a leaf
in the tree, we have
$$\sum_{{\textstyle J{\rm \ shallow}}} \Pr[J] \le k. \eqno(4.6)$$
This yields
$$\Pr[E^2_k] \le {1\over 3n-2k+1}\,{k\over 3n-2k+5}.$$
Combining this inequality with $(4.3)$ in $(4.2)$ completes the proof of Lemma 4.3.
\QED

\label{Proof of Lemma 4.4:}
From $(4.2)$, we have $\Pr[E_k] \ge \Pr[E^1_k]$, so it will suffice to obtain 
a lower bound to $\Pr[E^1_k]$.
Let us say that a node $K$ is a {\it strong precursor\/} to $E^1_k$ if
(1) every ancestor of $K$ (including $K$ itself) is single, and
(2) there are no reflectors at any ancestor of $K$, and
(3) ${\it length}(K) = k-1$, and
(4) $K$ is not a leaf.
At a strong precursor $K$ to $E^1_k$, there is exactly one child at which $E^1_k$ occurs.
Thus, for a strong precursor $K$ to $E^1_k$,
$$\Pr[E^1_k \mid K] = {1\over 3n-2{\it depth}(K)-1} \ge {1\over 3n}.$$
Let $A_k$ be the event that control reaches a strong precursor to $E^1_k$.
Since $\it length$ strictly increases along any path from the root to a leaf in the tree, at most one
node on any such path can be a strong precursor to $E^1_k$.
Thus
$$\Pr[A_k] = \sum_{{\textstyle K{\rm \ strong}\atop\textstyle{\rm precursor\ to\ }E^1_k}} \Pr[K].$$
It follows that
$$\eqalignno{
\Pr[E_k]
&\ge \Pr[E^1_k] \cr
& \cr
&\ge \sum_{{\textstyle K{\rm \ strong}\atop\textstyle{\rm precursor\ to\ }E^1_k}}
\Pr[K] \, \Pr[E^1_k \mid K] \cr
& \cr
&\ge {1\over 3n} 
\sum_{{\textstyle K{\rm \ strong}\atop\textstyle{\rm precursor\ to\ }E^1_k}}
\Pr[K] \cr
& \cr
&= {1\over 3n}\,\Pr[A_k]. &(4.7)\cr}$$
It remains to obtain a lower bound for $\Pr[A_k]$.

Let $B_k$ denote the event that control reaches a shallow double node $K$.
Let $C_k$ denote the event that some corner becomes a reflector at a shallow node $K$.
Let $F_k$ denote the event that the procedure terminates at a leaf $L$ with ${\it length}(L)$
less than $k$.
Let $G_k$ denote the event that control reaches a node $J$ with ${\it length}(J)$ at least $k$, 
without ever passing through a node $K$ with ${\it length}(K) = k-1$.
Then we have
$$\Pr[A_k] \ge 1 - \Pr[B_k] - \Pr[C_k] - \Pr[F_k, \overline{B_k}\,] 
- \Pr[G_k, \overline{B_k}, \overline{C_k}\,]. \eqno(4.8)$$
It remains to obtain upper bounds for the four probabilities on the right-hand-side.

First we deal with $\Pr[B_k]$.
If a node $K$ is double, but $K^*$ is single, we shall call $K^*$ a {\it double precursor}.
If a node $J$ is a double precursor, then 
(1) $J$ is single, so that ${\it head}(J)^- \not= {\it tail}(J)$, and
(2) $J$ has exactly one child of $J$ that is double (namely the one corresponding to drawing the corner
${\it head}(J)^-$ from the urn), and
(3) ${\it head}(J)^+$ leads to ${\it tail}(J)^+$.
Let $M^\alpha_J$ denote the event that at node $J$, control passes to a double child of $J$
by drawing corner $\alpha$ from the urn.
If $B_k$ occurs, then $M^\alpha_J$ must occur 
for some $\alpha$ at some shallow node $J$.
Thus
$$\Pr[B_k] \le
\sum_{{\textstyle J{\rm \ shallow}}}
\Pr[J] \sum_\alpha \Pr[M^\alpha_J \mid J].$$
Each term in the sum over $\alpha$ vanishes except possibly for the one with $\alpha = {\it head}(J)^-$,
so that we have
$$\sum_\alpha \Pr[M^\alpha_J \mid J] \le {1\over 3n-2{\it depth}(J)-1} \le {1\over 3n-2k+1}.$$
This yields
$$\Pr[B_k] \le {1\over 3n-2k+1}
\sum_{{\textstyle J{\rm \ shallow}}}
\Pr[J].$$
Using $(4.6)$, we obtain
$$\Pr[B_k] \le {k\over 3n-2k+1}, \eqno(4.9)$$
which is the desired upper bound for $\Pr[B_k]$.

Next we turn to $\Pr[C_k]$.
We have
$$\Pr[C_k] \le
\sum_{{\textstyle J{\rm \ shallow}}}
\Pr[J] \sum_\alpha \Pr[R^\alpha_J \mid J].$$
Using $(4.5)$, we have
$$\Pr[C_k] \le {1\over 3n-2k+1}
\sum_{{\textstyle J{\rm \ shallow}}}
\Pr[J],$$
and using $(4.6)$, we obtain
$$\Pr[C_k] \le {k\over 3n-2k+1}, \eqno(4.10)$$
which is the desired upper bound for $\Pr[C_k]$.

Next we deal with $\Pr[F_k, \overline{B_k}\,]$.
If $F_k$ occurs but $B_k$ does not, then $E^1_j$ must occur for some $1\le j\le k-1$.
Thus
$$\Pr[F_k, \overline{B_k}\,] \le \sum_{1\le j\le k-1} \Pr[E^1_j].$$
Using $(4.3)$, we obtain
$$\Pr[F_k, \overline{B_k}\,] \le {k\over 3n-2k+1},\eqno(4.11)$$
which is the desired upper bound for $\Pr[F_k, \overline{B_k}\,]$.

Finally, we turn to $\Pr[G_k, \overline{B_k}, \overline{C_k}\,]$.
At a given node $J$, a corner $\alpha$ will be called a {\it deflector\/} of size $l$
if $\alpha^-$ leads to some corner $\beta$ in $l$ steps.
Initially all deflectors are of size zero.
As links are added, deflectors may be extended, merged into marked or unmarked paths
or destroyed by being closed into unmarked cycles.
A deflector $\alpha$ of size $l$ will be called an {\it exit\/} from $J$ if 
(1) ${\it length}(J) \le k-2$, and (2) ${\it length}(J) + l + 1\ge k$.
If $G_k$ occurs but $B_k$ and $C_k$ do not, then at some shallow node $J$
the corner drawn from the urn must be an exit from $J$.
Thus
$$\Pr[G_k, \overline{B_k}, \overline{C_k}\,] \le
\sum_{{\textstyle J{\rm \ shallow}}}
\Pr[J]
\sum_{{\textstyle{\rm exit\ }\alpha\atop\textstyle{\rm from\ }J}}
\Pr[J^\alpha \mid J],$$
where $J^\alpha$ denotes the child of $J$ reached by drawing $\alpha$ from the  urn at $J$.
Since $J$ is shallow, we have
$$\Pr[j^\alpha \mid J] \le {1\over 3n-2k+1}.$$
This yields
$$\Pr[G_k, \overline{B_k}, \overline{C_k}\,] \le {1\over 3n-2k+1}
\sum_{{\textstyle J{\rm \ shallow}}}
\Pr[J] \, X_J,$$
where $X_J$ denotes the number of exits from $J$.
If $\pi$ is a path from the root to a leaf in the tree, we shall let
$$Y_\pi = \sum_{\textstyle{\rm shallow\ } J{\rm \ on\ }\pi} X_J$$
denote the number of exits from shallow nodes on $\pi$.
Then we have
$$\sum_{{\textstyle J{\rm \ shallow}}} \Pr[J]\, X_J = 
\sum_\pi \Pr[\pi] \, Y_\pi,$$
where the sum on the right-hand side is over all paths $\pi$ from the root to a leaf in the tree,
and $\Pr[\pi]$ is the probability that control follows the path $\pi$. 
Thus
$$\Pr[G_k, \overline{B_k}, \overline{C_k}\,] \le {1\over 3n-2k+1}
\sum_\pi \Pr[\pi] \, Y_\pi. \eqno(4.12)$$
We shall show that 
$$Y_\pi \le k \eqno(4.13)$$
for every path $\pi$.
To do this, we shall charge each exit from a shallow node $J$ on $\pi$
against an unmarked link that is added at a shallow node on $\pi$,
in such a way that at most one exit is charged against any link.
Since there are at most $k-1$ shallow nodes on $\pi$, each of which adds at most one unmarked link,
this will prove (4.13).

Suppose that corner $\alpha$ is an exit from node $J$ on $\pi$.
Let $s(\alpha, J)$ denote the size of the deflector $\alpha$ at $J$.
Since $\alpha$ is an exit from $J$, we must have
$${\it length}(J) + s(\alpha, J) + 1 \ge k,$$
so that the path from $\alpha^-$ contains at least $k - 1 - {\it length}(J)\ge 1$ links.
We shall charge the exit $\alpha$ from $J$ against the 
$\(k - 1 - {\it length}(J)\)$-th link (counting from the rear) in the path from $\alpha^-$.
It remains to verify that at most one exit from $\pi$ is charged against any link.
Suppose that exit $\alpha$ from node $J$ is the first exit from $\pi$ charged against link $\Lambda$.
Any exit $\beta$ from a later node $K$ on $\pi$ that is charged against a link on the path
containing $\alpha^-$ will be charged against a link that appears behind $\Lambda$ on this path
(since ${\it length}(K) \ge {\it length}(J)$ and any links in the path from $\alpha^-$ will also appear in
the path from $\beta^-$).
This completes the proof of $(4.13)$.

From $(4.12)$ and $(4.13)$ we have
$$\Pr[G_k, \overline{B_k}, \overline{C_k}\,] \le {k\over 3n-2k+1}
\sum_\pi \Pr[\pi].$$
Since 
$$\sum_\pi \Pr[\pi] \le 1,$$
we obtain
$$\Pr[G_k, \overline{B_k}, \overline{C_k}\,] \le {k\over 3n-2k+1}, \eqno(4.14)$$
which is the desired upper bound for $\Pr[G_k, \overline{B_k}, \overline{C_k}\,]$.

Substituting $(4.9)$, $(4.10)$, $(4.11)$ and $(4.14)$ into $(4.8)$, yields
$$\Pr[A_k] \ge 1 - {4k\over 3n-2k+1},$$
and substituting this into $(4.7)$ completes the proof of Lemma 4.4.
\QED
\sk

\heading{5. The Variance of the Number of Cycles}

\label{Theorem 5.1:}
$$\Var[h] = O(\log n).$$

\label{Proof:}
The proof is similar to that in Section 4,
so we shall merely sketch it.
We have
$$\Var[h] = \Ex[h^2] - \Ex[h]^2,$$
and we have seen that
$$\Ex[h] = \log n + O(1).$$
Thus to prove the theorem it will suffice to show that
$$\Ex[h^2] = (\log n)^2 + O(\log n). \eqno(5.1)$$

We have
$$\Ex[h^2] = \sum_{1\le k\le 3n} \; \sum_{1\le k'\le 3n}
\Ex[h_k \cdot h_{k'}],$$
where $h_k$ denotes the number of cycles of length $k$.

Our first step is to reduce the range of the summations over
$k$ and $k'$.
Since there are at most $5$ cycles of length exceeding $n/2$, we have
$$\eqalign{
\sum_{1\le k\le 3n} \; \sum_{1\le k'\le 3n}
\Ex[h_k \cdot h_{k'}] &\le
\sum_{1\le k\le n/2} \; \sum_{1\le k'\le n/2}
\Ex[h_k \cdot h_{k'}] + 10\Ex[h] + 25 \cr
&= \sum_{1\le k\le n/2} \; \sum_{1\le k'\le n/2}
\Ex[h_k \cdot h_{k'}] + O(\log n). \cr}$$
Thus to prove (5.1) it will suffice to show that
$$\sum_{1\le k\le n/2} \sum_{1\le k'\le n/2}
\Ex[h_k \cdot h_{k'}] = (\log n)^2 +O(\log n). \eqno(5.2)$$

We have
$$\sum_{1\le k\le n/2} \; \sum_{1\le k'\le n/2}
\Ex[h_k \cdot h_{k'}] = 
\sum_c \sum_{c'} 
\sum_{1\le k\le n/2} ; \sum_{1\le k'\le n/2}
{\Pr[c \incyc k, c' \incyc k']\over k\cdot k'},$$
where the outer sums are over all corners and
the event ``corner $c$ is in a cycle of length $k$''
has been abbreviated ``$c\incyc k$''.

If $c$ and $c'$ are corners of the same triangle, we shall write
$c\sim c'$, otherwise $c\not\sim c'$.
We have
$$\eqalign{
\sum_c \sum_{c' \sim c} \;
\sum_{1\le k\le n/2} \; &\sum_{1\le k'\le n/2}
{\Pr[c \incyc k, c' \incyc k']\over k\cdot k'} \cr
&=
\sum_c \sum_{c' \sim c} \;
\sum_{1\le k\le n/2} \; \sum_{1\le k'\le n/2}
{\Pr[c \incyc k] \, \Pr[c' \incyc k' \mid c\incyc k]\over k\cdot k'} \cr
&\le
\sum_c \sum_{c' \sim c} \;
\sum_{1\le k\le n/2} \; \sum_{1\le k'\le n/2}
{\Pr[c \incyc k] \, \Pr[c' \incyc k' \mid c\incyc k]\over k} \cr
&\le
\sum_c \sum_{c' \sim c} \;
\sum_{1\le k\le n/2} 
{\Pr[c \incyc k] \over k} \cr
&\le
3\sum_c 
\sum_{1\le k\le n/2} 
{\Pr[c \incyc k] \over k} \cr
&\le 3\,\Ex[h] \cr
&= O(\log n), \cr
}$$
since $\sum_{1\le k'\le n/2} \Pr[c' \incyc k' \mid c\incyc k] \le 1$,
and for any corner $c$, there are just $3$ corners $c'$ such that
$c\sim c'$.
Thus to prove (5.2) it will suffice to show that
$$\sum_c \sum_{c' \not\sim c} \;
\sum_{1\le k\le n/2} \; \sum_{1\le k'\le n/2}
{\Pr[c \incyc k, c' \incyc k']\over k\cdot k'} 
= (\log n)^2 + O(\log n). \eqno(5.3)$$

The probability distribution is invariant under a group of symmetries
that includes permutations of the triangles and cyclic permutations of
the corners of each triangle.
Since this group acts transitively on the $9n(n-1)$ pairs $(c,c')$ of
corners such that $c\not\sim c'$, we have
$$\eqalign{
\sum_c \sum_{c' \not\sim c} \;
\sum_{1\le k\le n/2} \; &\sum_{1\le k'\le n/2}
{\Pr[c \incyc k, c' \incyc k']\over k\cdot k'} \cr
&= 
9n(n-1)
\sum_{1\le k\le n/2} \; \sum_{1\le k'\le n/2}
{\Pr[c \incyc k, c' \incyc k']\over k\cdot k'}, \cr}$$
where $(c,c')$ is an arbitrary pair of corners such that $c\not\sim c'$.
Since $k$ and $k'$ now appear symmetrically in the last sum, we have
$$\eqalign{
9n(n-1)
\sum_{1\le k\le n/2} \; &\sum_{1\le k'\le n/2}
{\Pr[c \incyc k, c' \incyc k']\over k\cdot k'} \cr
&\le
18n(n-1)
\sum_{1\le k\le n/2} \; \sum_{k\le k'\le n/2}
{\Pr[c \incyc k, c' \incyc k']\over k\cdot k'}. \cr}$$
Thus to prove (5.3) it will suffice to show that
$$18n(n-1)
\sum_{1\le k\le n/2} \; \sum_{k\le k'\le n/2}
{\Pr[c \incyc k, c' \incyc k']\over k\cdot k'} = 
(\log n)^2 + O(\log n). \eqno(5.4)$$

Now we shall express the event ``corner $c$ is in a cycle of length $k$
and corner $c'$ is in a cycle of length $k'$'' as the disjoint union
to two events: 
``corner $c$ is in a cycle of length $k$
and corner $c'$ is in a disjoint cycle of length $k'$'',
which will be abbreviated ``$c\incyc k, c'\indisjcyc k'$'', and
 ``corner $c$ is in a cycle of length $k$
and corner $c'$ is in the same cycle of length $k'$'',
which can occur only if $k=k'$ and which will be abbreviated
``$c\incyc k, c'\insamecyc k$''.
We may also refer to the events 
``$c' \indisjcyc k'$'' and
``$c'\insamecyc k$''
conditioned on the event ``$c\incyc k$''.
We have
$$\eqalign{
18n(n-1)
\sum_{1\le k\le n/2} \; &\sum_{k\le k'\le n/2}
{\Pr[c \incyc k, c' \incyc k']\over k\cdot k'} \cr
&= 18n(n-1)
\sum_{1\le k\le n/2} \; \sum_{k\le k'\le n/2}
{\Pr[c \incyc k, c' \indisjcyc k']\over k\cdot k'} \cr
&\qquad+ 
18n(n-1)
\sum_{1\le k\le n/2} 
{\Pr[c \incyc k, c' \insamecyc k]\over k^2}. \cr
}$$
For the last sum we have
$$\eqalign{
18n(n-1)
&\sum_{1\le k\le n/2} 
{\Pr[c \incyc k, c' \insamecyc k]\over k^2} \cr
&=
18n(n-1)
\sum_{1\le k\le n/2} 
{\Pr[c \incyc k] \, \Pr[c' \insamecyc k \mid c\incyc k]\over k^2} \cr
&\le
6n
\sum_{1\le k\le n/2} 
{\Pr[c \incyc k] \over k} \cr
&\le 2\Ex[h] \cr
&= O(\log n), \cr
}$$
since $\Pr[c' \insamecyc k \mid c\incyc k]\le k/3(n-1)$
(there are at most $k-1$ corners of the cycle containing $c$
on different triangles from $c$, and the probability that 
a particular corner $c'$ of
the $3(n-1)$ corners on triangles different from $c$ is among them is 
thus at most $(k-1)/3(n-1) \le k/3(n-1)$).
Thus to prove (5.4) it will suffice to show that
$$18n(n-1)
\sum_{1\le k\le n/2} \; \sum_{k\le k'\le n/2}
{\Pr[c \incyc k, c' \indisjcyc k']\over k\cdot k'}
= (\log n)^2 + O(\log n). \eqno(5.5)$$

We have
$$
\eqalignno{
18n&(n-1)
\sum_{1\le k\le n/2} \; \sum_{k\le k'\le n/2}
{\Pr[c \incyc k, c' \indisjcyc k']\over k\cdot k'} \cr
&= 6n \sum_{1\le k\le n/2} {\Pr[c\incyc k]\over k} \;
3(n-1)  \sum_{k\le k'\le n/2} 
{\Pr[c'\indisjcyc k' \mid c\incyc k]\over k'}. &(5.6)\cr}$$
We can estimate $\Pr[c'\indisjcyc k' \mid c\incyc k]$  the same way as
we estimated $\Pr[c\incyc k]$ in Lemma 4.3 of the deriviation of 
$\Ex[h]$.
The only difference is that we start with a situation in which 
up to $k$ ribbons may already have been installed.
Thus we must replace $k$ by $k+k' \le 2k'$ in Lemma 4.3:
$$\Pr[c'\indisjcyc k' \mid c\incyc k]
\le {1\over 3n-4k'+1}
\left(1 + {2k'\over 3n-4k'+5}\right).$$
Thus for the inner sum in (5.6) we have
$$\eqalign{
3(n-1)  &\sum_{k\le k'\le n/2} 
{\Pr[c'\indisjcyc k' \mid c\incyc k]\over k'} \cr
&\le 3(n-1)  \sum_{k\le k'\le n/2}
{1\over k'}\cdot{1\over 3n-4k'+1}
\left(1 + {2k'\over 3n-4k'+5}\right) \cr
&\le \sum_{k\le k'\le n/2}
\left({1\over k'} + {4\over 3n-4k'+1}\right)
 \left(1 + {2k'\over 3n-4k'+5}\right) \cr
&= \log n - \log k + O(1). \cr
}$$
Furthermore, we still have
$$\Pr[c\incyc k] \le {1\over 3n-2k+1}\left(1 + {k\over 3n-2k+5}\right)$$
from Lemma 4.3.
Thus for the outer sum in (5.6) we have
$$\eqalign{
 6n &\sum_{1\le k\le n/2} {\Pr[c\incyc k]\over k}
\bigg(\log n - \log k + O(1)\bigg) \cr
&\le
6n \sum_{1\le k\le n/2}
{1\over k}\cdot{1\over 3n-2k+1}\left(1 + {k\over 3n-2k+5}\right) 
\bigg(\log n - \log k + O(1)\bigg) \cr
&\le2\sum_{1\le k\le n/2}
\left({1\over k} + {2\over 3n-2k+5}\right)
\left(1 + {k\over 3n-2k+5}\right) 
\bigg(\log n - \log k + O(1)\bigg) \cr
&= (\log n)^2 + O(\log n), \cr
}$$
where the main contribution comes from
$$2\sum_{1\le k\le n/2} {\log n - \log k \over k} = 
(\log n)^2 + O(\log n).$$
This proves (5.5) and thus completes the proof of the theorem.
\QED
\sk

\heading{6. The Probability of a Single Cycle}

We shall be concerned in this section with determining the probability
$$q_{n/2}  = \Pr[ h = 1]$$
that the fat-graph constructed from $n$ triangles contains a single boundary cycle.

\label{Lemma 6.1:}
We have
$$q_{2s} = 0$$
for $s\ge 0$.

\label{Proof:}
Since $h = n/2 + \chi$ and $\chi$ is even, $h$ has the same parity as $n/2$.
Thus $\Pr[h=1] = 0$ when $n/2$ is even.
\QED

Our main result is the following.

\label{Theorem 6.2:}
We have
$$q_{2s+1} = {2^{4s+1} \; 3^{3s+1} \; (4s+2)! \; (6s+2)! \; (6s+3)!
\over  (s+1)! \; (3s+1)! \; (12s+6)!}$$
for $s\ge 0$.

\label{Proof:}
Let $t = 2s+1 = n/2$.
Then if $h=1$, the single cycle has length $6t = 3n$.
We employ a technique attributed 
by Bessis, Itzykson and Zuber [12]
to J.~M. Drouffe.
This proof is uses the irreducible
characters $\chi^\alpha$ of the symmetric group $S_{6t}$,
indexed by the partitions $\alpha$ of $6t$
(that is, non-decreasing sequences of positive integers summing to $6t$).
The facts we need are found in the book by Sagan [46].

Let
$$Q_t = q_t \, (6t-1)!!$$
denote the number of pairings that result in 
a single cycle of length $6t$.
Then
$$Q_t = \sum_{\tau} 
\delta_{[\tau],[2^{3t}]} \delta_{[\sigma\tau],[6t]}, \eqno(6.1)$$
where the sum is over permutations $\tau$ in $S_{6t}$ and 
$\sigma$ is a fixed permutation of congugacy class $[3^{2t}]$
(that is, the conjugacy class of permutations containing $2t$ cycles of length $3$).
From the completeness relation for the characters of the symmetric 
group $S_{6t}$
(see Theorem 1.10.3 in Sagan [46]), we have
$$\sum_\alpha \chi^\alpha(\phi)\,\chi^\alpha(\psi) = 
{\delta_{[\phi],[\psi]} \, (6t)! \over \#[\phi]},$$
where the sum is over the partitions $\alpha$ of $6t$, $\phi$ and $\psi$ are permutations,
$[\phi]$ denotes the congugacy class containing $\phi$, and $\#[\phi]$ denotes its cardinatlity.
Substituting this relation in (6.1) yields
$$Q_t = {\#[2^{3t}] \, \#[6t] \over (6t)!^2}
\sum_{\alpha,\beta} \chi^\alpha([2^{3t}]) \, \chi^\beta([6t])
\sum_{\tau} \chi^\alpha(\tau) \, \chi^\beta(\sigma\tau).  \eqno(6.2)$$
We shall need the formula 
$$\sum_\tau \chi^\alpha(\tau) \, \chi^\beta(\sigma\tau) = 
{\delta_{\alpha,\beta} \, \chi^\alpha(\sigma) \, (6t)! \over \chi^\alpha([1^{6t}])}. \eqno(6.3)$$
This variant of the orthogonality relation for characters can be proved as follows.
Let $\{a_{i,j}\}_{1\le i,j\le v}$ and $\{b_{p,q}\}_{1\le p,q\le w}$
be the matrix elements of the representations corresponding to 
$\chi^\alpha$ and $\chi^\beta$, respectively.
From the proof of Theorem 1.9.3 in Sagan [46], we have
$$\sum_\tau a_{i,j}(\tau) \, b_{p,q}(\tau) = 
{(6t)! \, \delta_{\alpha,\beta} \, \delta_{i,q} \, \delta_{j,p} 
\over \chi^\alpha([1^{6t}])},$$ 
since $v = \chi^\alpha([1^{6t}])$.
(Note that $a_{i,j}(\tau^{-1}) = a_{i,j}(\tau)$, since every element is
conjugate to its inverse in the symmetric group.)
Setting $j=i$ and multiplying by $b_{q,p}(\sigma)$ yields
$$\sum_\tau a_{i,i}(\tau) \, b_{q,p}(\sigma) \, b_{p,q}(\tau) = 
{(6t)! \, \delta_{\alpha,\beta} \, \delta_{i,q} \, \delta_{i,p} \, a_{q,p}(\sigma)
\over  \chi^\alpha([1^{6t}])}.$$ 
Summing over $p$ yields
$$\sum_\tau a_{i,i}(\tau) \, b_{q,q}(\sigma\tau)  = 
{(6t)! \, \delta_{\alpha,\beta} \, \delta_{i,q} \, a_{q,i}(\sigma)
\over  \chi^\alpha([1^{6t}])};$$
summing over $q$ yields
$$\sum_\tau a_{i,i}(\tau) \,\chi^\beta(\sigma\tau)  = 
{(6t)! \, \delta_{\alpha,\beta} \, a_{i,i}(\sigma)\over  \chi^\alpha([1^{6t}])};$$
and summing over $i$ yields (6.3).
Substituting (6.3) in (6.2) yields
$$Q_t = {\#[2^{3t}] \, \#[6t] \over (6t)!}
\sum_\alpha {\chi^\alpha([2^{3t}]) \, \chi^\alpha([6t]) \, \chi^\alpha([3^{2t}])
\over \chi^\alpha([1^{6t}])}, \eqno(6.4)$$
since $\sigma\in [3^{2t}]$.

To evaluate the sum in (6.4), we observe that 
$$\chi^\alpha([6t]) = \cases{
(-1)^p, &if $\alpha = [6t-p, 1^p]$; \cr
& \cr
0, &otherwise. \cr
}$$
(This is Lemma 4.10.3 in Sagan [46].)
Thus, setting $\lambda(p) = [6t-p, 1^p]$, we have
$$Q_t = {\#[2^{3t}] \, \#[6t] \over (6t)!}
\sum_{0\le p\le 6t} {(-1)^p \,\chi^{\lambda(p)}([2^{3t}]) \, 
\chi^{\lambda(p)}([3^{2t}])
\over \chi^{\lambda(p)}([1^{6t}])}. \eqno(6.5)$$
The denominator $\chi^{\lambda(p)}([1^{6t}])$ is the dimension
of the representation coresponding to the partition $\lambda(p)$:
$$\chi^{\lambda(p)}([1^{6t}]) = {6t-1\choose p}.$$
(This follows from the Hook Formula, Theorem 3.10.2 in Sagan [46].)
The characters in the numerator are given by
$$\chi^{\lambda(p)}([2^{3t}]) = (-1)^{p+\lfloor p/2\rfloor}
{3t-1\choose \lfloor p/2\rfloor}$$
and
$$\chi^{\lambda(p)}([3^{2t}]) = (-1)^{p+\lfloor p/3\rfloor}
{2t-1\choose \lfloor p/3\rfloor}.$$
(These follow from the Murnaghan-Nakayama Rule, Theorem 4.10.2 in Sagan [46].)
Substituting these relations in (6.5) yields
$$Q_t = {\#[2^{3t}] \, \#[6t] \over (6t)!}
\sum_{0\le p\le 6t} {(-1)^{p+\lfloor p/2\rfloor+\lfloor p/3\rfloor}
\,{3t-1\choose \lfloor p/2\rfloor}
\, {2t-1\choose \lfloor p/3\rfloor}
\over {6t-1\choose p}}.$$
Finally, we have $\#[2^{3t}] = (6t-1)!!$ 
and $\#[6t] = (6t-1)!$, which
yields
$$Q_t = {(6t-1)!! \over 6t}
\sum_{0\le p\le 6t} {(-1)^{p+\lfloor p/2\rfloor+\lfloor p/3\rfloor}
\,{3t-1\choose \lfloor p/2\rfloor}
\, {2t-1\choose \lfloor p/3\rfloor}
\over {6t-1\choose p}}. \eqno(6.6)$$

The sign pattern in (6.6) has a period of $12$; this suggests that we
aggregate the terms in groups of $6$.
Thus we make the substitution $q = \lfloor p/6\rfloor$;
the result is
$$\eqalign{
Q_t &= {(6t-1)!! \over 6t} \times \cr
&\qquad \sum_{0\le q\le t-1}
{(-1)^q
\,{3t-1\choose 3q}
\, {2t-1\choose 2q}
\over {6t-1\choose 6q}}
\left[1 - {4\,(6q+1)\over (6t-6q-1)} 
+ {(6q+1)(6q+5)\over (6t-6q-1)(6t-6q-5)}\right]. \cr
}$$
Making the substitution $t = 2s+1$, we obtain
$$\eqalign{
Q_{2s+1} &= {(12s+5)!! \over 12s+6} \times \cr
& \sum_{0\le q\le 2s}
{(-1)^q
\,{6s+2\choose 3q}
\, {4s+1\choose 2q}
\over {12s+5\choose 6q}}
\left[1 - {4\,(6q+1)\over (12s-6q+5)} 
+ {(6q+1)(6q+5)\over (12s-6q+5)(12s-6q+1)}\right]. \cr
}$$
Thus to prove the theorem, we must show that
$$T(s) = \sum_{0\le q\le 2s} F(s,q) = 1, \eqno(6.7)$$
where
$$F(s,q) = {
(-1)^q \, (1 + 36q^2 + 4s - 72qs + 24s^2) \,
{4s+1 \choose 2q} \, {6s+2 \choose 3q} \,
(s+1)! \, (3s+1)! \, (12s+5)!
\over
2^{4s} \, 3^{3s} \, 
(1-6q+12s) \, (5-6q+12s) \,
{12s+5 \choose 6q} \,
(4s+2)! \, (6s+2)! \, (6s+3)!
}.$$
Define
$$G(s,q) = F(s,q) \, R(s,q),$$
where
$$R(s,q) = {A(s,q) \over B(s,q)},$$
$$\eqalign{
A(s,q) &= 
q \, (12s - 6q + 5) \, (12s - 6q + 1) \, \times \cr
&\qquad \(36q(1440s^5 + 5328s^4 + 6656s^3 + 2560s^2 - 734s - 527) - \cr
&\qquad\qquad 36q^2(1488s^4 + 4832s^3 + 5444s^2 + 2370s + 271) - \cr
&\qquad\qquad 1296q^4(2s +  1)\,(s+1) + 10368q^q(2s+1)\,(s+1)^2 - \cr
&\qquad\qquad 13824s^6 - 47232 s^5 - 27984 s^4 + 76288 s^3 + 75909 s +
15050\)\cr }$$
and
$$\eqalign{
B(s,q) &= 
2^4 \, 3^2 \,
(2s-q+1) \, (2s-q+2) \, (s+1) \,  \times \cr
&\qquad (2s+3)^2 \,(6s+5) \, (6s+7) \,
(24s^2 - 72qs + 4s + 36q^2 + 1). \cr
}$$
Then we have
$$F(s+1,q) - F(s,q) = G(s,q+1) - G(s,q).$$
Summing this result over all $q$ yields
$$T(s+1) - T(s) = 0,$$
since $F(s,q)$ vanishes outside the range of summation in (6.7).
Since $T(0) = 1$,we obtain (6.7) for all $s\ge 0$ by induction.
\QED

\label{Corollary 6.3:}
We have
$$\Pr[h=1] = \cases{
0, &if $n/2$ even; \cr
& \cr
{\displaystyle 2\over \displaystyle 3n} + 
O\left({\displaystyle 1\over \displaystyle n^2}\right), &if $n/2$ odd. \cr
}$$

\label{Proof:}
Applying Stirling's asymptotic formula in the form
$$n! = {(2\pi n)^{1/2} n^n\over e^n}
\left(1 + O\left({1\over n}\right)\right)$$
to Theorem 6.2 yields 
$$q_{2s+1} = {1\over 6s} + O\left({1\over s^2}\right).$$
This relation together with Lemma 6.1 yields the corollary.
\QED
\sk

\heading{7.  The Classification of Cycles} 

In this section we shall introduce a classification of boundary cycles in
the fat graph model, based on their ``self-interactions''.
In this classification, each cycle is either 
``simple'' or ``complex'', and each simple cycle is 
of some finite order.
We shall give a heuristic argument that predicts the expected number
of simple cycles of each finite order.
In the following sections, we shall give rigourous confirmations of these
predictions for the two lowest orders.

Let $C$ be a boundary cycle in a fat graph.
The {\it self-interaction surface\/} $S(C)$ of $C$ is the union of all
ribbons traversed by $C$ in both directions
(that is, all ribbons with both links marked), together with all triangles
incident with these ribbons
(that is, all triangles visited more than once by $C$).

If $S(C)$ contains two ribbons incident with a given triangle, then it
also contains the third ribbon incident with that triangle.
Thus every triangle in $S(C)$ is incident with either one ribbon or three.
Since every ribbon is incident with two triangles, it follows that the
number of triangles in a connected component of $S(C)$ is even.

We shall say that a connected component of
$S(C)$ is {\it simple\/} if it is simply connected, and that it
is {\it complex\/} otherwise.
A simple cycle will be said to have {\it order\/} $b$ if every connected component of 
$S(C)$ contains at most $2b$ triangles.
Thus a cycle has order zero if it traverses no ribbon in both directions
(or equivalently, if it visits at most one corner of any triangle);
it has order one if it visits at most
two corners in any triangle.

The top pairing in Figure 1 yields two simple cycles of order zero and one of order one;
the middle pairing yields three simple cycles of order zero; 
and the bottom pairing yields complex cycle.

Let us now consider, in a heuristic way, the expected number of simple cycles 
of various finite orders.
Let the random variable $s_b$ denote the number of simple cycles of
order at most $b$. 
We start with cycles of order zero.
A cycle cannot have a doubly traversed ribbon unless it has length at
least four, and it {\it must\/} have a doubly traversed ribbon if its
length exceeds $n$.
Thus we expect most short cycles to have order zero, and most long cycles
to have higher order.
Consideration of the ``birthday effect'' 
(if there are $n$ days in a year, then among $n^{1/2}$ people in a room there is 
a significant probability that two have the same birthday)
leads one to anticipate that
the transition will occur for cycles of length around $n^{1/2}$.
From the results of Section 4, we have that the expected number of cycles
of length $k$ is about $1/k$.
Thus we anticipate that 
$\Ex[s_0] \approx \sum_{1\le k\le n^{1/2}} {1\over k} = {1\over 2}\log n +
O(1)$. 
We can refine this estimate in the following way.
Although all doubly traversed ribbons are excluded from cycles of order zero,
we shall focus our attention on ``minimal forbidden self-interactions'',
which are isolated doubly traversed ribbons (that is,
components of $S(C)$ containing exactly two triangles joined by a single edge).
We assume the number $f_0$ of isolated doubly traversed ribbons is approximately
Poisson distributed, so that $\Pr[f_0 = 0] \approx \exp -\Ex[f_0]$. 
We shall see in the next paragraph that we have $\Ex[f_0] \approx k^2/6n$.
This leads us to anticipate that 
$$\eqalign{
\Ex[s_0]
&\approx \sum_{1\le k\le n} {1\over k} \exp(-k^2/6n) \cr
&\approx {1\over 2}\log (6n) + {\gamma\over 2}. \cr
}$$
In Section 8 we shall confirm this prediction in the form
$$\Ex[s_0] = {1\over 2}\log (6n) + {\gamma\over 2} 
+ O\left({\log n\over n^{1/3}}\right).$$

Let us now justify the approximation $\Ex[f_0] \approx k^2/6n$.
We consider how the algorithm of Section 4 can doubly traverse
a ribbon between the arc $(c_1^-,c_1)$ of triangle $t_1$ and the arc
$(c_2^-,c_2)$ of triangle $t_2$
(without doubly traversing any other ribbon incident with $t_1$ or $t_2$).
This can happen if corner
$c_1^+$ is drawn from the urn at some draw $d_1$,
then corner
$c_2$ is drawn at the immediately following draw $d_1+1$,
and finally corner $c_2^+$ is drawn at some subsequent draw $d_2>d_1+1$.
The two triangles $t_1$ and $t_2$ and their 
corners $c_1$ and $c_2$ can be chosen in about $9n^2$ ways.
The two draws $d_1$ and $d_2$ can be chosen in about $k^2/2$ ways.
Each of the three draws occurs with probability about $1/3n$,
for an overall probability of about $1/27n^3$.
The product of these factors must first be multiplied by a factor of 
$2$, since the same ribbon is doubly traversed (with the traversals
occurring in the opposite order) if $c_2^+$ is drawn at $d_1$,
$c_1$ is drawn at $d_1+1$ and $c_1^+$ is drawn at $d_2$.
Finally, we must divide by a factor of $2$, since the same ribbon is
doubly traversed if $c_1$ of $t_1$ is exchanged with $c_2$ of $t_2$.
The product of all these factors is $k^2/6n$.

We now turn to cycles of order at most one.
Here the minimal forbidden self-interaction is an isolated triply visited triangle,
and reasoning by analogy with the birthday effect we anticipate that
these will begin to appear when $k$ is about $n^{2/3}$.
This gives the estimate
$\Ex[s_1] \approx \sum_{1\le k\le n^{2/3}} {1\over k} = {2\over 3}\log n +
O(1)$.
Again we can refine this result by estimating the expectation of the
number $f_1$ of isolated triply visited triangles.
In the following paragraph we shall argue that 
$\Ex[f_1] \approx k^3/27n^2$, which leads us to anticipate that
$$\eqalign{
\Ex[s_1]
&\approx \sum_{1\le k\le 2n} {1\over k} \exp(-k^3/27n^2) \cr
&\approx {2\over 3}\log n + \log 3 + {2\gamma\over 3}. \cr
}$$
In Section 9 we shall confirm this prediction in the form
$$\Ex[s_1] = {2\over 3}\log n + \log 3 + {2\gamma\over 3} 
+ O\left({(\log n)^7\over n^{1/8}}\right).$$

Let us now justify the approximation  $\Ex[f_1] \approx k^3/27n^2$.
We consider how the algorithm of Section 4 can triply visit triangle
$t_4$ by doubly traversing the ribbon 
joining its arc $(c_4^-, c_4)$ to the arc
$(c_1^-,c_1)$ of a triangle $t_1$, the ribbon 
joining its arc $(c_4^+, c_4^-)$ to the arc
$(c_2^-,c_2)$ of a triangle $t_2$, and the ribbon 
joining its arc $(c_4, c_4^+)$ to the arc
$(c_3^-,c_3)$ of a triangle $t_3$
(without doubly traversing any of the other ribbons incident with
$t_1$, $t_2$ or $t_3$).
This can happen if corner
$c_1^+$ is drawn from the urn at some draw $d_1$,
corner
$c_4$ is drawn at the immediately following draw $d_1+1$,
corner
$c_2$ is drawn at the immediately following draw $d_1+2$,
then
corner $c_2^+$ is drawn at some subsequent draw $d_2>d_1+2$ and
corner
$c_3$ is drawn at the immediately following draw $d_2+1$,
and finally corner $c_3^+$ is drawn at some subsequent draw $d_3>d_2+1$.

The four triangles $t_1$, $t_2$, $t_3$ and $t_4$ and their 
corners $c_1$, $c_2$, $c_3$ and $c_4$ can be chosen in about $81n^4$ ways.
The three draws $d_1$, $d_2$ and $d_3$ can be chosen in about $k^3/6$
ways. 
Each of the six draws occurs with probability about $1/3n$,
for an overall probability of about $1/729n^6$.
The product of these factors must first be multiplied by a factor of 
$6$, since the same triangle is triply visited (with the visits
occurring in a different order) if the triangles $t_1$, $t_2$ and $t_3$
are visited in any of the five other permutations of order considered
above. 
Finally, we must divide by a factor of $3$, since the same triangle
is triply visited if $c_1$ of $t_1$, $c_2$ of $t_2$ and $c_3$ of $t_3$ are
cyclically permuted.
The product of all these factors is $k^3/27n^2$.

We now turn to cycles of order at most $b\ge 2$.
If the minimal forbidden self-interactions in the fat graph are shrunk to 
parts of the corresponding thin graph, the results are trees with
$2b+2$ vertices ($b+2$ leaves and $b$ internal vertices)
and $2b+1$ edges.
These trees are cubic plane trees, since every internal vertex has degree
$3$ and the trees are regarded as embedded in the plane,
so that two trees are regarded as isomorphic if there is a bijection
between their sets of vertices that preserves adjacency and also preserves
the cyclic order of the neighbours around each internal vertex.
(For each $b\in \{0,1,2,3\}$, there is just one such tree, 
up to isomorphism;
for each $b\ge 4$ there are more than one.)

We need to estimate the expectation of the number $f_b$ of occurrences of
such minimal forbidden self-interactions.
Consider a cubic plane tree with $2b$ vertices.
There are about $(3n)^{2b+2}$ ways of choosing $2b+2$ triangles and one corner
from each triangle.
The forbidden self-interaction will occur if $3b+3$ specific corners
are drawn in $b+2$ series of draws.
The probability of these corners being drawn is about $1/(3n)^{3b+3}$,
and the number of ways of choosing the $b+2$ draws that begin the 
successive series is about $k^{b+2}/(b+2)!$.
We must multiply by a factor of $(b+2)!$ accounting for the order in which
the visits to the self-interaction at the draws that begin the successive
series occur, and we must divide by a factor of $1$, $2$ or $3$
accounting for the automorphisms of the tree.
(A plane cubic tree can be symmetric about a vertex, with three
automorphisms, or symmetric about an edge, with two automorphisms,
or rigid, with just the trivial automorphism.)
Thus we have
$$\Ex[f_{b}] \approx {k^{b+2} \, T_{b} \over (3n)^{b+1}}, \eqno(7.1)$$
where $T_{b}$ is the sum over all cubic plane trees with $2b+2$ vertices
of the weight $1/q$ for each tree with $q$ automorphisms.

The eight trees corrresponding to minimal forbidden self-interactions
for $0\le b\le 4$, together with their weights, are shown in Figure 3.
The four trees on the left arise for $0\le b\le 3$; the four on the right
arise for $b=4$.

\label{Proposition 7.1:}
The sum of weights over plane cubic trees containing $2b+2$ vertices is
$$T_{b}  = {1\over (b+1)(b+2)}{2b\choose b}.$$

\label{Proof:}
We shall consider {\it rooted\/} trees, in which one vertex is
distinguished as the root.
We begin by considering cubic plane trees that are rooted at a leaf.
Let the generating function for such trees, in which the 
number of leaf-rooted cubic plane trees with $a$ vertices is 
the coefficient of $x^{a}$, be $F(x)$.
Such a tree either contains just two adjacent
vertices, or it can be constructed
by identifying the two roots of two such trees to form the neighbour of a
new leaf, which is the root of the constructed tree.
Thus we have $F(x) = x^2 + F(x)^2$, which implies
$$F(x) = {1 - (1 - 4x^2)^{1/2} \over 2}.$$

Let $\Phi_3(x)$ be the generating function for (unrooted) cubic plane trees
with three automorphisms.
Such a tree can be constructed by identifying the roots of three copies
of a leaf-rooted cubic plane tree to form an internal vertex 
(the vertex of symmetry) of the constructed tree.
Thus we have $\Phi_3(x) = F(x^3)/x^2$, which implies
$$\Phi_3(x) = {1 - (1 - 4x^6)^{1/2} \over 2x^2}.$$

Let $\Phi_2(x)$ be the generating function for (unrooted) cubic plane trees
with two automorphisms.
Such a tree can be constructed by identifying 
edges incident with the roots of two copies
of a leaf-rooted cubic plane tree to form an edge
(the edge of symmetry) of the constructed tree.
Thus we have $\Phi_2(x) = F(x^2)/x^2$, which implies
$$\Phi_2(x) = {1 - (1 - 4x^4)^{1/2} \over 2x^2}.$$

Let $G(x)$ be the generating function for rooted cubic plane trees
(where now the root may be any vertex, either leaf or internal).
Such a tree is either a leaf-rooted cubic plane tree, or it can be 
constructed by identifying the roots of three leaf-rooted cubic plane
trees to form an internal vertex of the constructed tree.
This procedure constructs each internally rooted cubic plane tree three
times, unless the three trees that are combined are isomorphic,
in which case the resulting tree, which has three automorphisms,
is constructed once.
Thus we have $G(x) = F(x) + F(x)^3/3x^2 + 2F(x^3)/3x^2$, which implies
$$G(x) = 
{3 - (1 + 2x^2)(1 - 4x^2)^{1/2} - 2(1 - 4x^6)^{1/2} \over 6x^2}.$$

Let $H_1(x)$ be the generating function for rooted rigid cubic plane trees
(where the root may be any vertex, and the trees have only the trivial
automorphism).
The number of rooted cubic plane trees with two automorphisms is 
given by the generating function
$$H_2(x) = y {d\over dy}{F(y)\over y}\bigg\vert_{y=x^2}
 = {1 - (1-4x^2)^{1/2} \over 2x^2 (1-4x^2)^{1/2}},$$
and the number of rooted cubic plane trees with three automorphisms is 
given by the generating function
$$H_3(x) = x {d\over dy}{F(y)}\bigg\vert_{y=x^3}
 = {2x^4\over (1-4x^6)^{1/2}}.$$
Thus we have
$$\eqalign{
H(x)
&= G(x) - H_2(x) - H_3(x) \cr
&= {6 - (1 + 2x^2)(1 - 4x^2)^{1/2} \over 6x^2}
 - {1\over 2x^2(1 - 4x^4)} 
- {1 + 2x^6 \over 3x^2(1 - 4x^6)^{1/2}}. \cr
}$$

Let $\Phi_1(x)$ be the generating function for (unrooted) rigid cubic plane
trees.
We have
$$\eqalign{
\Phi_1(x) 
&= \int_0^x {H_1(y) \, dy \over y} \cr
&= {(1-4x^2)^{3/2} + 3(1-4x^4)^{1/2} + 2(1-4x^6)^{1/2} - 6\over 12x^2}.
\cr }$$
The generating function $\Psi(x)$ for the numbers $T_b$ is then given by
$$\eqalign{
\Psi(x)
&= \left(\Phi_1(y) + {1\over 2}\Phi_2(y)x +
{1\over 3}\Phi_3(y)\right)\bigg\vert_{y = x^{1/2}} \cr
&= {(1-4x)^{3/2} - (1-6x)\over 12x}. \cr
\cr }$$
The proposition now follows using the binomial theorem.
\QED

The numbers $6T_{b}$ (which are integers)
occur as a generalization
of the Catalan numbers in the work of Gessel [47], who asked for an
enumerative interpretation of them.
Such an enumerative interpretation is provided by considering 
drawings of cubic plane trees in which every edge is drawn at an angle 
that is a multiple of $\pi/3$ from a reference line.
Cubic plane trees themselves are equivalent to ``flexagons'', which
have been counted by Oakley and Wisner [48].

Substituting the result of Proposition 7.1 into (7.1) yields
$$\Ex[f_b] \approx {k^{b+2} \,  \over (3n)^{b+1}} \cdot
{1\over (b+1)(b+2)}{2b\choose b}.$$
Thus we anticipate that 
$$\eqalign{
\Ex[s_b] 
&\approx \sum_{1\le k\le 3n} {1\over k} 
\exp -\left({k^{b+2} \,  \over (3n)^{b+1}} \cdot
{1\over (b+1)(b+2)}{2b\choose b}\right) \cr
&\approx {b+1\over b+2}\log (3n) + {(b+1)\gamma\over b+2}
- {1\over b+2}\log \left(
{1\over (b+1)(b+2)}{2b\choose b}\right). \cr
}$$
We note that
$${1\over b+2}\log \left(
{1\over (b+1)(b+2)}{2b\choose b}\right) \to \log 4$$
as $b\to\infty$.
This suggests the conjecture that the number of complex cycles is asymptotic to
$\log 4 = 1.386\ldots\,$.
\sk

\heading{8. Cycles of Order Zero}

A simple cycle of order zero has length at most $n$
(since each of $n$ triangles is visited at most once), and for $1\le k\le n$
the expected number of such cycles of length $k$ is
$${1\over k}\cdot{3^k \, n!\over (n-k)!}\cdot
{(3n-2k-1)!!\over(3n-1)!!}.$$ 
(The middle factor counts the ways of choosing an
ordered list of $k$ corners, no two on the same triangle.
The last factor gives the probability of 
connecting these corners into a cycle in the order given.
This counts each cycle $k$ times, hence the first factor.)
The expected number $s_0$ of simple cycles of order zero is thus
$$\Ex[s_0] = \sum_{1\le k\le n}
{1\over k}\cdot{3^k \, n!\over (n-k)!}\cdot
{(3n-2k-1)!!\over(3n-1)!!}.$$

\label{Theorem 8.1:}
We have
$$\Ex[s] = {1\over 2}\log (6n) + {\gamma\over 2} 
+ O\left({\log n\over n^{1/3}}\right).$$

\label{Proof:}
If we define
$$\eqalign{
F_k &= {\displaystyle\prod_{1\le j\le k} (3n - 3j + 3) 
\over \displaystyle\prod_{1\le j\le k} (3n - 2j + 1)} \cr
&= {\displaystyle\prod_{1\le j\le k} \left(1 - {j-1\over n}\right) 
\over \displaystyle\prod_{1\le j\le k} \left(1 - {2j-1\over 3n}\right)},
\cr }$$
we can write
$$\Ex[s] =  \sum_{1\le k\le n}
{1\over k} F_k.$$

For the numerator of $F_k$ we have
$$\eqalign{
\prod_{1\le j\le k} \left(1 - {j-1\over n}\right) 
&=
\exp \sum_{1\le j\le k} \log
\left(1 - {j\over n} + O\left({1\over n}\right)\right) \cr
&=
\exp \sum_{1\le j\le k} 
\left(-{j\over n} + O\left({1\over n}\right) 
+ O\left({j^2\over n^2}\right)\right) \cr 
&=
\exp 
\left(-{k^2\over 2n} + O\left({k\over n}\right) 
+ O\left({k^3\over n^2}\right)\right). \cr
}$$
Similar estimation of the denominator yields
$$\prod_{1\le j\le k} \left(1 - {2j-1\over 3n}\right)
= \exp 
\left(-{k^2\over 3n} + O\left({k\over n}\right) 
+ O\left({k^3\over n^2}\right)\right).$$
Thus we have
$$F_k = \exp
\left(-{k^2\over 6n} + O\left({k\over n}\right) 
+ O\left({k^3\over n^2}\right)\right).$$

Define
$$l = 
\lceil (6n\log n)^{1/2}\rceil.$$
Then we have
$$\eqalign{
F_l
&= O\left({1\over n}\right). \cr
}$$
Since $F_k$ is a decreasing function of $k$, we have
$$\eqalign{
\sum_{l<k\le n} {1\over k} \cdot F_k
&=O\left({1\over n}  \sum_{l<k\le n} {1\over k}\right) \cr
&= O\left({1\over l}\right) \cr
&= O\left({1\over n^{1/3}}\right). \cr
}$$
Thus to prove the theorem it will suffice to show that
$$\sum_{1\le k\le l} {1\over k} \cdot F_k = 
{1\over 2}\log (6n) + {\gamma\over 2} 
+ O\left({\log n\over n^{1/3}}\right). \eqno(8.1)$$

For $k\le l$ we have
$$\eqalign{
F_k
&= \exp
\left(-{k^2\over 6n}  
+ O\left({(\log n)^{3/2}\over n^{1/2}}\right)\right) \cr
&= \left(1 + O\left({(\log n)^{3/2}\over n^{1/2}}\right)\right)
\exp\left(-{k^2\over 6n}\right), \cr
}$$
so that
$$\sum_{1\le k\le l} {1\over k} \cdot F_k
= \left(1 + O\left({(\log n)^{3/2}\over n^{1/2}}\right)\right)
\sum_{1\le k\le l} {1\over k} \cdot 
\exp \left(-{k^2\over 6n}\right).$$
Thus to prove (8.1) it will suffice to show that
$$\sum_{1\le k\le l} {1\over k} \cdot 
\exp \left(-{k^2\over 6n}\right) = 
{1\over 2}\log (6n) + {\gamma\over 2} 
+ O\left({\log n\over n^{1/3}}\right). \eqno(8.2)$$

Define
$$m = \lceil (6n)^{1/3}\rceil.$$
Then for $k\le m$ we have
$$\eqalign{
F_k 
&= \exp \;O\left({1\over n^{1/3}}\right) \cr
&= 1 + O\left({1\over n^{1/3}}\right). \cr
}$$
This yields
$$\eqalign{
\sum_{1\le k\le m} {1\over k} \cdot 
\exp \left(-{k^2\over 6n}\right)
&=
\left(1 + O\left({1\over n^{1/3}}\right)\right)
\sum_{1\le k\le m} {1\over k} \cr
&=
\left(1 + O\left({1\over n^{1/3}}\right)\right)
\left({1\over 3}\log (6n) + \gamma 
+ O\left({1\over n^{1/3}}\right)\right), \cr 
}$$
where $\gamma = 0.5772\ldots$ is Euler's constant.
Thus to prove (8.2) it will suffice to show that
$$\sum_{m < k\le l} {1\over k} \cdot 
\exp \left(-{k^2\over 6n}\right) = 
{1\over 6}\log (6n) - {\gamma\over 2} 
+ O\left({\log n\over n^{1/3}}\right). \eqno(8.3)$$

We have
$$\sum_{m < k\le l} {1\over k} \cdot 
\exp \left(-{k^2\over 6n}\right) = 
\int_m^l {1\over z} \cdot 
\exp \left(-{z^2\over 6n}\right) \,dz + 
O\left({1\over n^{1/3}}\right),$$
since we may bound the difference between a sum and an integral by the
total variation of the integrand.
The substitution $z = (6ny)^{1/2}$ yields
$$
\int_m^l {1\over z} \cdot 
\exp \left(-{z^2\over 6n}\right) \,dz
=
{1\over 2}\int_{m^2/6n}^{l^2/6n} {1\over y} \cdot 
\exp(-y) \,dy.$$
Since
$$\eqalign{
\int_{l^2/6n}^\infty {1\over y} \cdot 
\exp(-y) \,dy 
&=
O\left( {6n\over l^2} \cdot 
\exp\left(-{l^2\over 6n}\right)\right) \cr
&= O\left({\log n\over n}\right), \cr
}$$
we may raise the upper bound of the integral from $l^2/6n$ to $\infty$,
and thus obtain an expression in terms of the exponential integral:
$$\eqalign{
{1\over 2}\int_{m^2/6n}^{l^2/6n} {1\over y} \cdot 
\exp(-y) \,dy
&=
{1\over 2}\int_{m^2/6n}^\infty {1\over y} \cdot 
\exp(-y) \,dy + O\left({\log n\over n}\right) \cr
&=
-{1\over 2}\Ei\left(-{m^2\over 6n}\right) 
+ O\left({\log n\over n}\right), \cr 
}$$
where
$$-\Ei(-x) = \int_x^\infty {1\over w}\cdot \exp(-w) \, dw$$
(see Lebedev [49], \S 1.3).
Using the asymptotic expansion
$$-\Ei(-x) = \log {1\over x} - \gamma + O(x),$$
we obtain
$$\eqalign{
-{1\over 2}\Ei\left(-{m^2\over 6n}\right)
&= {1\over 2}\log {6n\over m^2} - {\gamma\over 2} 
+ O\left({m^2\over 6n}\right) \cr 
&= {1\over 6}\log (6n) - {\gamma\over 2} 
+ O\left({1\over n^{1/3}}\right). \cr 
}$$
Combining these results yields (8.3), and thus completes the proof of the
theorem.
\QED
\sk

\heading{9. Cycles of Order  One}

We begin by deriving an exact formula for the expected number of simple cycles
of order one.

\label{Proposition 9.1:}
We have
$$\eqalignno{
\Ex[s_1] = \sum_{1\le k\le 2n} \; \sum_{0\le j\le k/4} \;
&{1\over k}
\left[{k-2j\choose 2j} + {k-2j-1\choose 2j-1}\right] 
\, (2j-1)!! \, \times\cr 
&\qquad{ 3^{k-2j} \, n! \over
(n-k+2j)!  }\cdot
{(3n-k+j-1)!! \over (3n-1)!!}. &(9.1)\cr
}$$

\label{Proof:}
If a cycle visits each of the $n$ triangles at most twice, it can have
length at most $2n$.
Furthermore, if a cycle visits each triangle at most twice, then each
traversal of a doubly traversed ribbon must be immediately followed by 
the traversal of a singly traversed ribbon, and thus such a cycle of
length $k$ can have at most $k/4$ doubly traversed ribbons.
Thus it remains to show that the summand 
in (9.1) is the expected number of 
cycles of length $k$ with $j$ doubly traversed ribbons.

Consider a cyclic directed graph $C = (V, E)$ with 
vertices $V = \{v_1, \ldots, v_k\}$ and
edges $E = \{(v_1,v_2), \ldots, (v_{k-1}, v_k), (v_k, v_1)\}$.
A {\it template\/} $T = (I,J)$ comprises a set $I\subseteq E$ of $2j$
edges of $C$, subject to the condition that no two edges in $I$ are
consecutive, together with a partition $J$ of $I$ into $j$ pairs of edges.
The number of templates is 
$$\left[{k-2j\choose 2j} + {k-2j-1\choose 2j-1}\right] 
\, (2j-1)!!,$$
since there are ${k-2j\choose 2j} + {k-2j-1\choose 2j-1}$ ways of
choosing the $2j$ edges in $I$ (there are 
${k-2j\choose 2j}$ ways that exclude the edge $(v_k, v_1)$, 
and ${k-2j-1\choose 2j-1}$ ways that include it), and $(2j-1)!!$
ways to partition these edges to form the $j$ pairs in $J$.
(We agree that $(-1)!! = 1$, as a special case of the formula
$(2j-1)!! = (2j)! / 2^j \, j!$.)

The cyclic group of order $k$ acts on $C$ in an obvious way, and this
induces an action on the set of templates.
Let $p(T)$ denote the number of templates in the orbit of $T$ under this
action, and let $q(T)$ denote the order of the automorphism group of $T$.
Then we have $p(T) \, q(T) = k$.

Next consider a boundary cycle in the fat graph with length $k$ and $j$
doubly traversed ribbons.
We shall construct a template as follows.
Pick a corner $c_1$ on the cycle, and then define $c_2, \ldots, c_k$ to
be the the successive corners of the cycle.
For each ribbon that is doubly traversed by the links
$(c_f, c_{f+1})$ and $(c_g, c_{g+1})$ of the cycle, put the edges
$(v_f, v_{f+1})$ and $(v_g, v_{g+1})$ into $I$ and put the pair
$\{(v_f, v_{f+1}), (v_g, v_{g+1})\}$ into $J$.
Different templates can be obtained from the same boundary cycle
by different choices of the corner $c_1$;
$p(T)$ different templates are obtained
in this way, since if we advance the choice
of $c_1$ by $p(T)$ corners we obtain the same template.

Now consider the boundary cycles corresponding 
to a given template $T$.
The cycle visits $k$ corners on $k-2j$ triangles, but once the first
visited corner on a triangle is chosen, the remaining visited corner on
that triangle is determined. 
The triangles can be chosen in $n! / (n-k+2j)!$
ways, and the first visited corner on each triangle can be chosen in
$3^{k-2j}$ ways.
Different choices of triangles and corners can produce the same
boundary cycle; each boundary cycle is produced $q(T)$
times, since if we advance the choices of vertices $k/q(T) = p(T)$
positions we produce the same cycle.
For such a cycle to occur
in the fat graph, $k-j$ specific draws must occur, corresponding to
the first traversals of the $k-j$ ribbons traversed by the cycle.
The probability of these draws occurring is 
$(3n-k+j-1)!! / (3n-1)!!$.

Finally, the expected number of boundary cycles of length $k$ with 
$j$ doubly traversed ribbons is
$$\sum_{T} {1\over p(T)}\cdot { 3^{k-2j} \, n! \over q(T) \,
(n-k+2j)!  }\cdot
{(3n-k+j-1)!! \over (3n-1)!!},$$
where the sum is over all templates $T$.
Since $p(T) \, q(T) = k$, we obtain the summand in (9.1).
\QED

In principle, the asymptotic behaviour of $\Ex[s_1]$ can be obtained by
direct analysis (9.1).
We have found it most convenient, however, to bound the terms with the
smallest and largest values of $k$ through separate arguments, and thus
to analyze (9.1) only for a set of intermediate values that includes the
transition (around $n^{2/3}$) from significant terms to negligible ones.
We give these separate arguments in the following two propositions.

\label{Proposition 9.2:}
Let $m = n^{5/8}$.
The expected number of cycles of order at most one and length at most
$m$ is
$$\log m + \gamma + O\left({1\over n^{1/8}}\right).$$

\label{Proof:}
From the estimates in the proof of Theorem 4.1, we have that the expected 
number of cycles of length at most $m$ is
$$\sum_{1\le k\le m} \left({1\over k} + O\left({1\over n}\right)\right)
= \log m + \gamma +  O\left({1\over n^{3/8}}\right).$$
The obstruction to a cycle having order at most one is a triply visited
triangle.
Thus it will suffice to show that the expected number of cycles of length 
at most $m$ that contain a triply visited triangle is 
$O(1/n^{1/8})$.
This expectation is in turn at most the expected number of 
triply visited triangles 
that are created during the first $m$ iterations of the
algorithm of Section 4.

There are three ways in which a triply visited triangle can occur:
the three ribbons incident with it can be incident with three other triangles,
or with just two, or with just one.
The expected number of triply visited triangles
for which the ribbons are incident with 
three other triangles was dealt with heuristically in Section 6.
This argument can be made rigourous, at the cost of lowering our sights to
an upper bound that holds within a constant factor. 
There are $O(n^4)$ ways of
choosing corners 
$c_1$, $c_2$, $c_3$ and $c_4$ of triangles $t_1$, $t_2$, $t_3$ and $t_4$;
there are $O(m^3)$ ways of choosing three draws $d_1$, $d_2$ and $d_3$;
and the probability that the corners drawn at draws
$d_1$, $d_1+1$, $d_1+2$, $d_2$, $d_2+1$ and $d_3$ create a
triply visited triangle is $O(1/n^6)$.
Thus this expectation is $O(m^3/n^2) = O(1/n^{1/8})$.
A similar argument shows that the case in which the three ribbons incident
with the triply visited triangle are incident with two other triangles
gives a smaller contribution, $O(n^4\,k^2/n^6) = O(m^2/n^2) = O(1/n^{3/4})$.
Finally, the case in which the three ribbons are all incident with one other
triangle (which can happen only if $k=6$) gives a still smaller
contribution, $O(n^2/n^3) = O(1/n)$.
\QED

For the next proposition, we shall need the following lemma.

\label{Lemma 9.3:}
Let $X$ denote the number of successes among $h$ trials that each succeed
independently with probability at most $p$.
Then
$$\Pr[X > 2hp] \le (e/4)^{hp}.$$
If on the other hand the trials succeed with probability at least
$q$, then 
$$\Pr[X < hq/2] \le (2/e)^{hq/2}.$$

\label{Proof:}
For the first inequality we may assume that 
the trials succeed with
probability exactly $p$, since the resulting random variable is majorized
by $X$.
We may also assume that
$p < 1/2$, since otherwise
$\Pr[X > 2hp] = 0$.
If 
$$Y = \cases{
0, &if $X\le 2hp$, \cr
1, &if $X > 2hp$, \cr}$$
then $\Pr[X > 2hp] = \Ex[Y]$.
If $Z = T^{X-2hp}$ (where $T > 1$ is a parameter to be chosen later),
then $Y\le Z$ and so $\Ex[Y] \le \Ex[Z]$.
Thus it will suffice to estimate $\Ex[Z]$.
Since $X$ is the sum of $h$ independent random variables that assume the 
value $1$ with probability $p$ and the value $0$ with probability $1-p$,
$T^X$ is the product of $h$ independent random variables that have 
expected value $pT + 1-p$. 
Thus
$$\Ex[Z] = (pT + 1-p)^h \, T^{-2hp}.$$
Choosing $T = 2(1-p)/(1-2p)$ and using the inequality $1+x \le e^x$
yields the first inequality of the lemma.
A similar argument yields the second inequality.
\QED

Lemma 9.3 is an instance of Chernoff's inequality.

\label{Proposition 9.4:}
Let $l = \lceil n^{2/3}\log n\rceil$.
Then the expected number of cycles of order at most one with length 
exceeding $12l$ is $O(1/n)$.

\label{Proof:}
From the  proof of Theorem 4.1, we have that the expected number
of cycles of order at most one with length exceeding $12l$ is
$$\eqalign{
3n \sum_{k > 12l} 
&{\Pr[\hbox{$c_0$ in a $k$-cycle of order at most one}] \over k} \cr
 &\le
(n/4l)
\Pr[\hbox{$c_0$ in a cycle of order at most one and length exceeding $12l$}]
\cr }.$$
Thus it will suffice to show that
$$
\Pr[\hbox{$c_0$ in a cycle of order at most one at length at least $12l$}]
 = O\left({1\over n^2}\right).$$
This probability is at most the probability that the algorithm of Section
4 runs for $6l$ iterations without creating a triply visited triangle.

To estimate this probabilty, we shall use an analysis that assigns
colours to the corners in the urn during the execution of the algorithm.
Corners will initially be white, but may 
subsequently be recoloured red or blue.
The rules for colouring will be such that if a blue corner is drawn
from the urn during the first $6l$ iterations, then either the cycle
containing $c_0$ has length less than $12l$ or it contains a
triply visited triangle.
Thus it will suffice to show that the probability that no blue corner is
drawn during the first $6l$ iterations is $O(1/n)$.

The rules for colouring are as follows.
Initially, all corners are white except for $c_0^-$, which is red,
and $c_0^+$, which is blue.
If ever a blue corner is drawn, we stop the analysis.
Whenever a white corner $c$ is drawn, then
if the immediately preceding corner drawn was white, 
we colour $c^+$ red, but 
if the immediately preceding corner drawn was red, 
we colour $c^+$ blue.
It is easy to verify that drawing a blue corner either closes the cycle
(which then has length at most $12l$) or creates a triply visited triangle.
(It is possible to create a triply visited triangle without drawing
a blue corner; the rules given have been chosen to be as simple as 
possible while yielding the desired upper bound.)

The rules specify how corners in the urn become coloured.
There are two ways in which coloured corners can leave the urn:
they can be drawn or they can be removed from the urn after becoming the 
{\it head\/} of the marked path.
But at any given iteration,
a corner $c$ can be removed in this way only if one particular
corner $d$ (such that $d^-$ is the rear of the unmarked path whose
front is $c$) is drawn.
Thus the probability of removing a coloured corner from the urn is never
greater than the probability of drawing it.

We shall now obtain estimates for the numbers of corners of various
colours in the urn at various times during the first $6l$ iterations.
We shall divide these  iterations into three {\it phases\/}
(phases I, II and III), each comprising $2l$ iterations.

At any time during these $6l$ iterations, at most $12l$ corners have been
drawn or removed from the urn, and thus the urn always contains at most $3n$
and at least $3n - 12l \ge 3n/2$ corners.
Since at most one corner is coloured per iteration, 
there are always at least $(3n-12l) - 6l = 3n-18l$ white corners in the urn,
and thus the probability
of drawing a white corner is always at least $(3n - 18l)/3n \ge 1/2$.

We start by considering the number of corners coloured red during phase I.
A corner is coloured red whenever two consecutive draws yield white
corners.
For each of $l$ disjoint pairs of consecutive draws, this probability
is at least $(1/2)(1/2) = 1/4$.
Using Lemma 9.3 with $h = l$ and $q = 1/4$, we have that, except with
probability at most $(2/e)^{l/8}$, at least $l/8$ corners are coloured red
during phase I.

Next we consider the number of red corners drawn or removed during phases
I and II.
During these $4l$ iterations, there are never more than $4l$ red corners
in the urn.
Thus the probability that a red corner is drawn or removed in any
iteration is at most $8l/3n$.
Using Lemma 9.3 with $h=4l$ and $p=8l/3n$, we have that, except with
probability at most $(e/4)^{32l^2/3n}$, at most $64l^2/3n$
red corners are drawn or removed during phases I and II.
Thus, except with probability at most
$$(2/e)^{l/8} + (e/4)^{32l^2/3n},$$
there are at least
$l/8 - 64l^2/3n \ge l/16$ red corners in the urn throughout phase II.

Next we consider the number of corners coloured blue during phase II.
A corner is coloured blue whenever two consecutive draws yield 
a red corner followed by a white corner.
For each of $l$ disjoint pairs of consecutive draws, this probability
is at least $\((l/16)/3n\)(1/2) = l/96n$.
Using Lemma 9.3 with $h = l$ and $q = l/96n$, we have that, except with
probability at most $(2/e)^{l^2/192n}$, at least $l^2/192n$ corners are
coloured blue during phase II.

We shall also need an upper bound, holding with high probability,
for the number of corners coloured blue during phases I, II and III.
For each of $6l$ draws, the probability
of colouring a corner blue is at most $\(6l\big/ (3n/2)\) = 4l/n$.
Using Lemma 9.3 with $h = 6l$ and $p = 4l/n$, we have that, except with
probability at most $(e/4)^{24l^2/n}$, at most $48l^2/n$ corners are
coloured blue during phases I, II and III.

Next we consider the number of blue corners drawn or removed during
phases I, II and III.
Assuming that there are never more than $48l^2/n$ blue corners
in the urn during these $6l$ iterations, we have that the 
probability that a blue corner is drawn or removed in any
iteration is at most $(48l^2/n)/(3n/2) = 32l^2/n^2$.
Using Lemma 9.3 with $h=6l$ and $p=32l^2/n^2$, we have that, except with
probability at most $(e/4)^{192l^3/n^2}$, at most $384l^3/n^2$
blue corners are drawn or removed during phases I, II and III.
Thus, except with probability at most
$$(2/e)^{l/8} + (e/4)^{32l^2/3n} + 
(2/e)^{l^2/192n} + (e/4)^{24l^2/n}
+ (e/4)^{192l^3/n^2},$$
there are at least
$l^2/192n - 384l^3/n^2 \ge l^2/384n$ blue corners in the urn throughout
phase III.

Finally
we consider the probability that no blue corner is drawn during phase III.
Assuming that there are at least $l^2/384n$ blue corners in the urn
throughout phase III, we have that the probability of drawing
a blue corner at each iteration is at least 
$(l^2/384n)/3n = l^2/1152n^2$, and thus that
the probability of not drawing a blue corner in any of these $2l$
iterations is at most
$(1 - l^2/1152n^2)^{2l} \le (1/e)^{l^3/576n^2}$.
Thus, except with probability at most
$$(2/e)^{l/8} + (e/4)^{32l^2/3n} + 
(2/e)^{l^2/192n} + (e/4)^{24l^2/n}
+ (e/4)^{384l^3/n^2} + (1/e)^{l^3/576n^2},$$ 
either the cycle  closes 
or a triply visited triangle is created before the length of the
cycle reaches $12l$.
These six contributions are all 
$$\exp -\Omega\((\log n)^3\) = O\left({1\over n^2}\right).$$
\QED

\label{Theorem 9.5:}
The expected number of cycles of order at most one is
$$\Ex[s_1] = {2\over 3}\log n + \log 3 + {2\gamma\over 3} + 
O\left({(\log n)^7\over n^{1/8}}\right).$$

\label{Proof:}
Using Propositions 9.1, 9.22 and 9.4, it will suffice to show that
$$\eqalignno{
\sum_{m\le k\le 12l} \; &{1\over k} \; \sum_{0\le j\le k/4}
\left[{k-2j\choose 2j} + {k-2j-1\choose 2j-1}\right] 
\, (2j-1)!! \, \times\cr 
&{ 3^{k-2j} \, n! \over
(n-k+2j)!  }\cdot
{(3n-k+j-1)!! \over (3n-1)!!} = 
{1\over 3}\log {27n^2\over m^3}  - {\gamma\over 3} + 
O\left({(\log n)^7\over n^{1/8}}\right), &(2)\cr
}$$
where $m = n^{5/8}$ and $l = \lceil n^{2/3} \log n\rceil$.
To do this, we shall use a ``bootstrapping'' technique, whereby crude
estimates are used to reduce the range of the summation over $j$, so that
more delicate estimates will be applicable over the reduced range.
Specifically, we shall show that for certain terms the inner summand in (9.2)
is $O(1/n^3)$.
Since there are $O(n^2)$ such terms, this will imply that the 
total contribution of these terms is $O(1/n)$, and thus can be neglected.

First, we shall show that all term with $j\ge 6a$, where
$$a = {k^2\over 6n},$$ 
may be neglected.
Using the estimates
$${k-2j\choose 2j} + {k-2j-1\choose 2j-1}
\le 2\,{k-2j\choose 2j}
\le {2\,k^{2j}\over (2j)!}, \eqno(9.3)$$
$$(2j-1)!! 
= {(2j)!\over 2^j \, j!}
\le (2j)! \, \left({e\over 2j}\right)^j, \eqno(9.4)$$
$${3^{k-2j} \, n! \over (n-k+2j)!}
\le (3n)^{k-2j},$$
and
$$\eqalignno{
{(3n-k+j-1)!! \over (3n-1)!!}
&\le {1\over (3n)^{k-j}}
\exp \left({k^2\over 3n} + O\left({k\over n}\right)
+ O\left({k^3\over n^2}\right)\right) \cr
&\le {1\over (3n)^{k-j}}
\exp \left({2a} + O\left((\log n)^3\right)\right), &(9.5)\cr
}$$
we have that the inner summand in (9.2) is at most
$$\left({ea\over j}\right)^j \; 
\exp\left({2a} + O\left((\log n)^3\right)\right).$$
This is a decreasing function of $j$ for $j\ge a$,
and for $j=6a$ it is equal to
$$\left({e^{4/3}\over 6}\right)^{6a} \; 
\exp O\left((\log n)^3\right)
= 
\exp\left(- \Omega(n^{1/4}) + O\left((\log n)^3\right)\right),$$
since $k\ge m = \Omega(n^{5/8})$.
Thus all terms with $j\ge 6a$ are $O(1/n^3)$ and may therefore be
neglected. 
Thus it will suffice to show that
$$\eqalignno{
\sum_{m\le k\le 12l} \; &{1\over k} \; \sum_{0\le j\le 6a}
\left[{k-2j\choose 2j} + {k-2j-1\choose 2j-1}\right] 
\, (2j-1)!! \, \times\cr 
&{ 3^{k-2j} \, n! \over
(n-k+2j)!  }\cdot
{(3n-k+j-1)!! \over (3n-1)!!} = 
{1\over 3}\log {27n^2\over m^3}  - {\gamma\over 3} + 
O\left({(\log n)^7\over n^{1/8}}\right). &(9.6)\cr
}$$

Next, we shall show that all terms with $j\le a-d$ or $j\ge a+d$, where
$$d = a^{1/2}(\log n)^2,$$
may be neglected.
Using the estimates (9.3), (9.4),
$$\eqalign{
{3^{k-2j} \, n! \over (n-k+2j)!}
&\le (3n)^{k-2j} \,
\exp \left( - {(k-2j)^2\over 2n} + O\left({k\over n}\right)
+ O\left({k^3\over n^2}\right)\right) \cr
&\le (3n)^{k-2j} \,
\exp \left( - {k^2\over 2n}  + {2kj\over n} + O\left({k\over n}\right)
+ O\left({k^3\over n^2}\right)\right) \cr
&\le (3n)^{k-2j} \,
\exp \left( - {3a} 
+ O\left((\log n)^3\right)\right) \cr
}$$
(where we have used $j\le 6a$) and (9.5), we have that the inner summand 
in (9.6) is at most
$$\eqalign{
\left({ea\over j}\right)^j \; \exp\left(- a + O\left((\log n)^3\right)\right)
&=
\exp\left(a\left({j\over a} - {j\over a}\log {j\over a} - 1\right)
+  O\left((\log n)^3\right)\right)\cr 
&\le
\exp\left( - {(j-a)^2\over 6a} +  O\left((\log n)^3\right)\right),\cr
}$$
since $x - x\log x - 1 \le -(x-1)^2/6$ for $0\le x\le 6$ and $j\le 6a$.
Thus if $j\le a-d$ or $j\ge a+d$, we have that the inner summand 
in (9.6) is at most
$$\exp\left(-{d^2\over 6a} +  O\left((\log n)^3\right)\right)
= \exp\left(-\Omega\((\log n)^4\) +  O\left((\log n)^3\right)\right).$$
Thus all terms with $j\le a-d$ or $j\ge a+d$ are $O(1/n^3)$ and may therefore
be neglected. 
Thus it will suffice to show that
$$\eqalignno{
\sum_{m\le k\le 12l} \; &{1\over k} \; \sum_{a-d\le j\le a+d}
\left[{k-2j\choose 2j} + {k-2j-1\choose 2j-1}\right] 
\, (2j-1)!! \, \times\cr 
&{ 3^{k-2j} \, n! \over
(n-k+2j)!  }\cdot
{(3n-k+j-1)!! \over (3n-1)!!} = 
{1\over 3}\log {27n^2\over m^3}  - {\gamma\over 3} + 
O\left({(\log n)^7\over n^{1/8}}\right). &(9.7)\cr
}$$

Finally, we have the estimates
$$\eqalign{
{k-2j\choose 2j} + {k-2j-1\choose 2j-1}
&=
{k-2j\choose 2j} \exp \; O\left({j\over k}\right) \cr
&= {k^{2j}\over (2j)!}
\exp \left( -{6j^2\over k} + 
O\left({j\over k}\right) + O\left({j^3\over k^2}\right)\right) \cr
&= {k^{2j}\over (2j)!}
\exp \left( -{k^3\over 6n^2} + 
O\left({(\log n)^4\over n^{1/6}}\right) \right) \cr
}$$
(where we have used $j = a + O(d)$ and $k = O(l)$),
$$\eqalign{
(2j-1)!!
&= {(2j)!\over 2^j \, j!} \cr
&= (2j)! \, \left({1\over 2\pi j}\right)^{1/2}
\left({e\over 2j}\right)^j \, \exp \; O\left({1\over j}\right) \cr
&= (2j)! \, \left({1\over 2\pi a}\right)^{1/2}
\left({e\over 2j}\right)^j \, 
\exp \; O\left({d\over a}\right) \cr
&= (2j)! \, \left({1\over 2\pi a}\right)^{1/2}
\left({e\over 2j}\right)^j \, 
\exp \; O\left({(\log n)^2\over n^{1/8}}\right) \cr
}$$
(where we have used $j = a + O(d)$ and $k = \Omega(m)$),
$$\eqalign{
&{3^{k-2j} \, n! \over (n-k+2j)!} \cr
&\qquad =
(3n)^{k-2j} \, 
\exp \left( - {(k-2j)^2\over 2n} + O\left({k\over n}\right)
- {(k-2j)^3\over 6n^2} + O\left({k^2\over n^2}\right)
+ O\left({k^4\over n^3}\right)
\right) \cr
&\qquad =
(3n)^{k-2j} \, 
\exp \left( - {k^2\over 2n} + {2kj\over n}
- {k^3\over 6n^2} 
+ O\left({k^4\over n^3}\right)
\right) \cr
&\qquad =
(3n)^{k-2j} \, 
\exp \left( - {k^2\over 2n} 
+ {k^3\over 6n^2} 
+ O\left({(\log n)^4\over n^{1/6}}\right)
\right) \cr
}$$
(where we have used $j = a + O(d)$ and $k = O(l)$) and
$$\eqalign{
&{(3n-k+j-1)!! \over (3n-1)!!} \cr
&\qquad = {1\over (3n)^{k-j}} \,
\exp \left({(k-j)^2\over 3n} + O\left({k\over n}\right)
+ {2(k-j)^3\over 27n^2} + O\left({k^2\over n^2}\right)
+ O\left({k^4\over n^3}\right)
\right) \cr
&\qquad = {1\over (3n)^{k-j}} \,
\exp \left({k^2\over 3n} - {2kj\over 3n}
+ {2k^3\over 27n^2} 
+ O\left({k^4\over n^3}\right)
\right) \cr
&\qquad = {1\over (3n)^{k-j}} \,
\exp \left({k^2\over 3n} 
- {k^3\over 27n^2} 
+ O\left({(\log n)^4\over n^{1/6}}\right)
\right) \cr
}$$
(where we have used $j = a + O(d)$ and $k = O(l)$).
Thus the inner summand in (9.7) is 
$$\eqalign{
\left({1\over 2\pi a}\right)^{1/2} \, &\left({ea\over j}\right)^j
\, \exp
\left(-a - {k^3\over 27n^2} 
+ O\left({(\log n)^2\over n^{1/8}}\right)\right) \cr
&= \left({1\over 2\pi a}\right)^{1/2} \, \exp
\left(-{(j-a)^2\over 2a} + O\left({\vert j-a\vert^3\over a^2}\right)
- {k^3\over 27n^2}  + O\left({(\log n)^2\over n^{1/8}}\right)\right) \cr
&= \left({1\over 2\pi a}\right)^{1/2} \, \exp
\left(-{(j-a)^2\over 2a} 
- {k^3\over 27n^2}  + O\left({(\log n)^6\over n^{1/8}}\right)\right), \cr
}$$
since $x - x\log x - 1 = (x-1)^2/2 + O\(\vert x-1\vert^3\)$, $j = a + O(d)$
and $k\ge m = \Omega(n^{5/8})$.
Thus the inner sum in (9.7) is
$$\left({1\over 2\pi a}\right)^{1/2} \,
\exp\left( - {k^3\over 27n^2} + O\left({(\log n)^6\over n^{1/8}}\right)\right)
\; \sum_{a-d\le j\le a+d} \exp -{(j-a)^2\over 2a}.$$
We have
$$\eqalign{
\sum_{a-d\le j\le a+d} \exp -{(j-a)^2\over 2a}
&= \sum_{-d\le i\le d} \exp -{i^2\over 2a} \cr
&= \int_{-d}^d \exp -{\xi^2\over 2a} \, d\xi + O(1), \cr
}$$
since the error in estimating a sum by the corresponding integral is at most
the total variation of the summand.
Furthermore, we have
$$\eqalign{
\int_{-d}^d \exp -{\xi^2\over 2a} \, d\xi
&= a^{1/2} \int_{-d/a^{1/2}}^{d/a^{1/2}} \exp -{\eta^2\over 2} \, d\eta \cr
&= a^{1/2} \int_{-\infty}^\infty \exp -{\eta^2\over 2} \, d\eta 
+ \exp -\Omega\((\log n)^4\) \cr
&= (2\pi a)^{1/2} + \exp -\Omega\((\log n)^4\), \cr
}$$
since $\int_{-\infty}^{-\beta} \exp-{\eta^2\over 2}\,d\eta = 
\int_\beta^\infty \exp-{\eta^2\over 2}\,d\eta \le
{1\over\beta}\exp-{\beta^2\over 2}$, $d/a^{1/2} = (\log n)^2$
and $\int_{-\infty}^\infty \exp-{\eta^2\over 2}\,d\eta = (2\pi)^{1/2}$.
Thus the inner sum in (9.7) is
$$\exp\left( - {k^3\over 27n^2} 
+ O\left({(\log n)^6\over n^{1/8}}\right)\right).$$
Summing this result over $m\le k\le 12l$ yields
$$\exp \; O\left({(\log n)^6\over n^{1/8}}\right) \;
\sum_{m\le k\le 12l} {1\over k} \, \exp - {k^3\over 27n^2}. \eqno(9.8)$$
Evaluating this sum in the same way as the one in Section 8, we have
$$\eqalign{
\sum_{m\le k\le 12l} {1\over k} \, \exp - {k^3\over 27n^2}
&= \int_m^{12l} {1\over \xi} \, \exp - {\xi^3\over 27n^2} \, d\xi
+O\left({1\over n^{5/8}}\right) \cr
&= {1\over 3}\int_{m^3/27n^2}^{(12l)^3/27n^2} {1\over \eta} \, 
\exp -\eta
\, d\eta +O\left({1\over n^{5/8}}\right) \cr
&= {1\over 3}\log {27n^2\over m^3} - {\gamma\over 3} + 
O\left({1\over n^{1/8}}\right), \cr
}$$
since $\int_\alpha^\infty {1\over\eta}\exp-\eta\,d\eta = \log{1\over \alpha} - \gamma
+O(\alpha)$ for $\alpha\to 0$ and 
$\int_\beta^\infty {1\over\eta}\exp-\eta\,d\eta \le {1\over \beta}\exp-\beta$ for
$\beta\to \infty$.
Substituting these results into (9.8) yields (9.7)
\QED
\sk

\heading{10. Conclusion}

The principal problem left open by the present work is to determine completely
the probability distribution for the random variable $h$.
If, for example, this were done by giving the generating function for $h$,
one could presumably obtain the moments by differentiation (as was done
for $h'$ and $h''$ in Section 2), and thus confirm or refute the conjectured
asymptotics of $\Ex[h]$ and $\Var[h]$.
The most promising approaches to this problem appear to lie in the connection
with matrix models.
An analogous model with a quartic interaction has been described by 
Bessis, Itzykson and Zuber [12], who relate the coefficients 
in a conjectured asymptotic expansion
to the enumeration of certain graphs according to their genera.
The existence of this asymptotic expansion and the interpretation of the coefficients
has been rigourously established by Ercolini and McLaughlin [50].
Their approach does not, however, appear to provide a 
derivation of the results of this paper, let alone our more precise conjectures.
Another approach would be to extend Harer and Zagier's [43] analysis of 
glueings of the $(2n)$-gon to $n$ triangles
(see also Penner [44] and Itzykson and Zuber [45]).
This approach leads naturally to the $\xi$-fold integral
$$\int_{-\infty}^{+\infty} \cdots \int_{-\infty}^{+\infty}
\left(\sum_{1\le i\le \xi} x_i^3\right)^n \,
e^{-{1\over 2}\sum_{1\le i\le \xi} x_i^2}
\prod_{1\le i<j\le \xi} (x_i - x_j)^2 
\, dx_1 \cdots dx_\xi.$$
If this integral could be evaluated as an analytic function of $\xi$, the result
would be (apart from an easily calculable normalizing factor) the 
desired generating function for $h$.

Another open problem is to determine the asymptotic behaviour of the expected number of
complex boundary cycles in the fat-graph model.
The heuristic analysis of Section 7 suggests that this number tends to a constant,
$\log 4$, but we cannot rigourously exclude either that it grows unboundedly 
or tends to zero.
Yet another problem is to analyze the effect of dropping that requirement that all surfaces be orientable;
this could be done by allowing each glueing to occur in either of two ways with equal probability.

Finally, our model assigns equal probabilities to all possible glueings of the
triangles, with the result that the most likely outcome is a connected surface
of high genus.
It would be of interest to explore the consequences of departing 
from this assumption.
One could, for example, 
associate a self-interaction energy with cycles that doubly traverse
ribbons.
Such an interaction that penalized double traversals of ribbons would presumably 
tend to increase the total Euler characteristic, increasing the number of components
and decreasing their genera, with results that would be more similar to 
matrix models in the large-$N$ limit.
\sk

\heading{11. Acknowlegdment}

The rational function $R(s,q)$ certifying the identity in Section 6 was found using 
Marko Petkovsek's 
program {\tt Zeil} in the Mathematica package {\tt gosper.m} available
at {\tt http://www.cis.upenn.edu/~wilf/AeqB.html}.
\vfill\eject

\heading{12. References}

\refinbook 1; J. A. Wheeler;
``
Geometrodynamics and the Issue of the Final State'';
in: B.~S. and C.~M. DeWitt (Ed's);
Relativit\'e, Groupes et Topologie;
Gordon and Breach, New York, 1964, pp.~315--520.

\ref 2; S. W. Hawking;
``Spacetime Foam'';
Nuclear Phys.\ B; 144 (1978) 349--362.

\refinbook 3; S. W. Hawking;
``The Path-Integral Approach to Quantum Gravity'';
in: S.~W. Hawking and W.~Israel (Ed's);
General Relativity: An Einstein Centenary Survey;
Cambridge University Press, New York, 1979, pp.~759--785.

\ref 4; S. Carlip;
``Dominant Topologies in Euclidean Quantum Gravity'';
Classical and Quantum Gravity; 15 (1998) 2629--2638,
[{\tt arXiv:gr-qc/9710114}].

\ref 5; S. Carlip;
``Spacetime Foam and the Cosmological Constant'';
Phys.\ Rev.\ Lett; 79 (1997) 4071--4074,
[{\tt arXiv:gr-qc/9708026}].

\ref 6; A. A. Markov;
``The Problem of Homeomorphy'' (Russian);
Proc.\ Internat.\ Congress Mathematicians;
Cambridge University Press, London, 1960, pp.~300--306.

\ref 7; W. W. Boone, W. Haken and V. Poenaru;
``On Recursively Unsolvable Problems in Topology and Their  
Classification'';
Contributions to Mathematical Logic;
North-Holland, Amsterdam, 1968, pp.~37--74.

\ref 8; K. Schleich and D. M. Witt;
``Generalized Sums over Histories for Quantum Gravity. 
II. Simplicial Conifolds'';
Nuclear Phys.\ B; 402 (1993) 469--528,
[{\tt arXiv:gr-qc/9307019}].

\ref 9; G. 't Hooft;
``A Planar Diagram Theory for Strong Interactions'';
Nuclear Physics B; 72 (1974) 461--473.

\ref 10; G. 't Hooft;
``A Two-Dimensional Model for Mesons'';
Nuclear Phys.\ B; 75 (1974) 461--470.

\ref 11; E.~Brezin, C.~Itzykson, G.~Parisi and J.~B.~Zuber;
``Planar Diagrams'';
Commun.\ Math.\ Phys.; 59 (1978) 35--51.

\ref 12; D.~Bessis, C.~Itzykson and J.~B.~Zuber;
``Quantum Field Theory Techniques In Graphical Enumeration'';
Adv.\ Appl.\ Math.;  1  (1980) 109--157.

\ref 13; E.~Brezin and V.~A.~Kazakov;
``Exactly Solvable Field Theories of Closed Strings'';
Phys.\ Lett.\ B; 236 (1990) 144--150.

\ref 14; M.~R.~Douglas and S.~H.~Shenker;
``Strings in Less than One-Dimension'';
Nucl.\ Phys.\ B; 335  (1990) 635--654.

\ref 15; D.~J.~Gross and A.~A.~Migdal;
``A Nonperturbative Treatment of Two-Dimensional Quantum Gravity'';
Nucl.\ Phys.\ B; 340  (1990) 333--365.

\ref 16; D.~J.~Gross and A.~A.~Migdal;
``Nonperturbative Two-Dimensional Quantum Gravity'';
Phys.\ Rev.\ Lett.; 64  (1990) 127--130.

\ref 17; P. Di Francesco, P. Ginsparg and J. Zinn-Justin;
``2-D Gravity and Random Matrices'';
Phys.\ Rep.; 254 (1995) 1--133.

\ref 18; A. M. Polyakov;
``Quantum Geometry Of Bosonic Strings'';
Phys.\ Lett.\ B;
103 (1981) 207--210.

\ref 19; O. Alvarez;
``Theory of Strings with Boundaries: Fluctuations, Topology, and 
Quantum Geometry'';
Nuclear Phys.\ B;
216 (1983) 125--184.

\refbook 20; J. Polchinski;
String Theory; 
Cambridge University Press, New York, 1998, v.~1, pp.~86--90.

\ref 21; E. Witten;
``Noncommutative Geometry and String Field Theory'';
Nuclear Phys.\ B; 268 (1986) 253--294.

\ref 22; G. T. Horowitz, J. Lykken, R. Rohm and A. Strominger'';
``A Purely Cubic Interaction for String Field Theory'';
Phys.\ Rev.\ Lett.; 57 (1986) 283--286.

\item{23} H. Ooguri and C. Vafa.
``The $C$-Deformation of Gluino and Non-planar Diagrams'',
{\tt ArXiv:hep-th/0302109}.

\ref 24; T.~Regge;
``General Relativity without Coordinates'';
Nuovo Cim.; 19  (1961) 558--571.

\ref 25; R.~M.~Williams and P.~A.~Tuckey;
``Regge calculus: A Bibliography and Brief Review'';
Class.\ Quant.\ Grav.; 9  (1992) 1409--1422.

\ref 26; R.~M.~Williams;
``Recent progress in Regge calculus'';
Nucl.\ Phys.\ Proc.\ Suppl.; 57  (1997) 73
[{\tt arXiv:gr-qc/9702006}].

\refbook 27; J. Ambj\o rn, M. Carfora and A. Marzuoli; 
The Geometry of  Dynamical Triangulations;
Springer-Verlag, New York, 1997.

\ref 28; D.~Weingarten;
``Euclidean Quantum Gravity on a Lattice'';
Nucl.\ Phys.\ B; 210  (1982) 229--245.

\ref 29; J.~Ambj\o rn, B.~Durhuus and J.~Frohlich;
``Diseases of Triangulated Random Surface Models, and Possible Cures'';
Nucl.\ Phys.\ B; 257  (1985) 433--449.

\refinbook 30; J.~Frohlich;
``Survey of Random Surface Theory'';
in: J.~Ambj\o rn, B.~J. Durhuus and J.~L. Petersen (Ed's);
Recent Developments in Quantum Field Theory;
North-Holland, Amsterdam, 1985, pp.~67--93.

\ref 31; F.~David;
``Randomly Triangulated Surfaces in $-2$ Dimensions'';
Phys.\ Lett.\ B; 159  (1985) 303--306.

\ref 32; F.~David;
``A Model of Random Surfaces with Nontrivial Critical Behavior'';
Nucl.\ Phys.\ B; 257  (1985) 543--576.

\ref 33; V.~A.~Kazakov, A.~A.~Migdal and I.~K.~Kostov;
``Critical Properties of Randomly Triangulated Planar Random Surfaces'';
Phys.\ Lett.\ B; 157  (1985) 295--300.

\refinbook 34; G. Ponzano and T. Regge;
``Semiclassical Limit of Racah Coefficients'';
in: F.~Bloch (Ed.);
Spectroscopic and Group Theoretical Methods in Physics;
North-Holland, New York, 1968.

\refinbook 35; R. Penrose; 
``Angular Momentum: An Approach to Combinatorial Space-Time''; 
in: T.~Bastin (Ed.);
Quantum Theory and Beyond; 
Cambridge University Press,
Cambridge, UK, 1971, pp.~151-180. 

\refinbook 36; R. Penrose; 
``On the Nature of Quantum Geometry''; 
in:  J.~Klauder;
Magic without Magic; 
Freeman, San Francisco, 1972, pp.~333-354.

\refinbook 37; R. Penrose; 
Combinatorial Quantum Theory and Quantized Directions''; 
in:  L.~Hughston and R.~Ward (Ed's);  
Advances in Twistor Theory; 
Pitman Advanced Publishing Program, San Francisco, 1979, pp.~301-317.

\ref 38; J. Barrett and L. Crane;
``Relativistic Spin Networks and Quantum Gravity''; 
Jour.\ Math.\ Phys.; 39 (1998) 3296--3302.

\refinbook 39; J. Baez; 
``Strings, Loops, Knots and Gauge Fields''; 
in: J.~Baez (Ed.);
 Knots and Quantum Gravity; 
Oxford University Press, Oxford, 1994, pp.~133--168.

\ref 40; J.~C.~Baez;
``An Introduction to Spin Foam Models of BF Theory and Quantum Gravity'';
Lect.\ Notes Phys.;  543 (2000) 25--93,  [{\tt arXiv:gr-qc/9905087}].

\ref 41; J. Hartle;
``Unruly Topologies in $2$-d Quantum Gravity'';
Class.\ Quantum Grav.; 2 (1985) 707--720.

\ref 42; L. Heffter;
``\"Uber das Problem der Nachbargebiete'';
Math.\ Ann.; 38 (1891) 477--508.

\ref 43; J. Harer and D. Zagier;
``The Euler Characteristic of the Moduli Space of Curves'';
Invent.\ Math.; 85 (1986) 457--485.

\ref 44; R. C. Penner;
``Perturbative Series and the Moduli Space of Riemann Surfaces'';
J. Differential Geometry; 27 (1988) 35--53.

\ref 45; C. Itzykson and J.-B. Zuber;
``Matrix Integration and Combinatorics of Modular Groups'';
Commun.\ Math.\ Phys.; 134 (1990) 197--207.

\refbook 46; B. Sagan;
The Symmetric Group:  Representations, Combinatorial Algorithms,
and Symmetric Functions;
Springer-Verlag, New York, 2001.

\ref 47; I. M. Gessel;
``Super Ballot Numbers'';
J. Symbolic Comp.; 14 (1992) 179--194.

\ref 48; C. O. Oakley and R. J. Wisner;
``Flexagons'';
Amer.\ Math.\ Monthly; 64 (1957) 143--154.

\refbook 49; N. N. Lebedev;
Special Functions and Their Applications;
Prentice-Hall, Englewood Cliffs, NJ, 1965.

\ref 50; N. M. Ercolani and K. D. T.-R. McLaughlin;
``Asymptotics of the Partition Function for Random Matrices via
Riemann-Hilbert Techniques, and Applications to Graphical Enumeration'';
Internat.\ Math.\ Research Notices; 2003 (2003) 755-820,
[{\tt http://xxx.lanl.gov/math-ph/0211022}].

\vfill\eject
\input epsf
\hsize=160 true mm
\vsize=250 true mm

\nopagenumbers

\midinsert
\vskip .1in
\hskip 0in
\epsfxsize=5.2in
\epsfysize=6.8in
\epsfbox{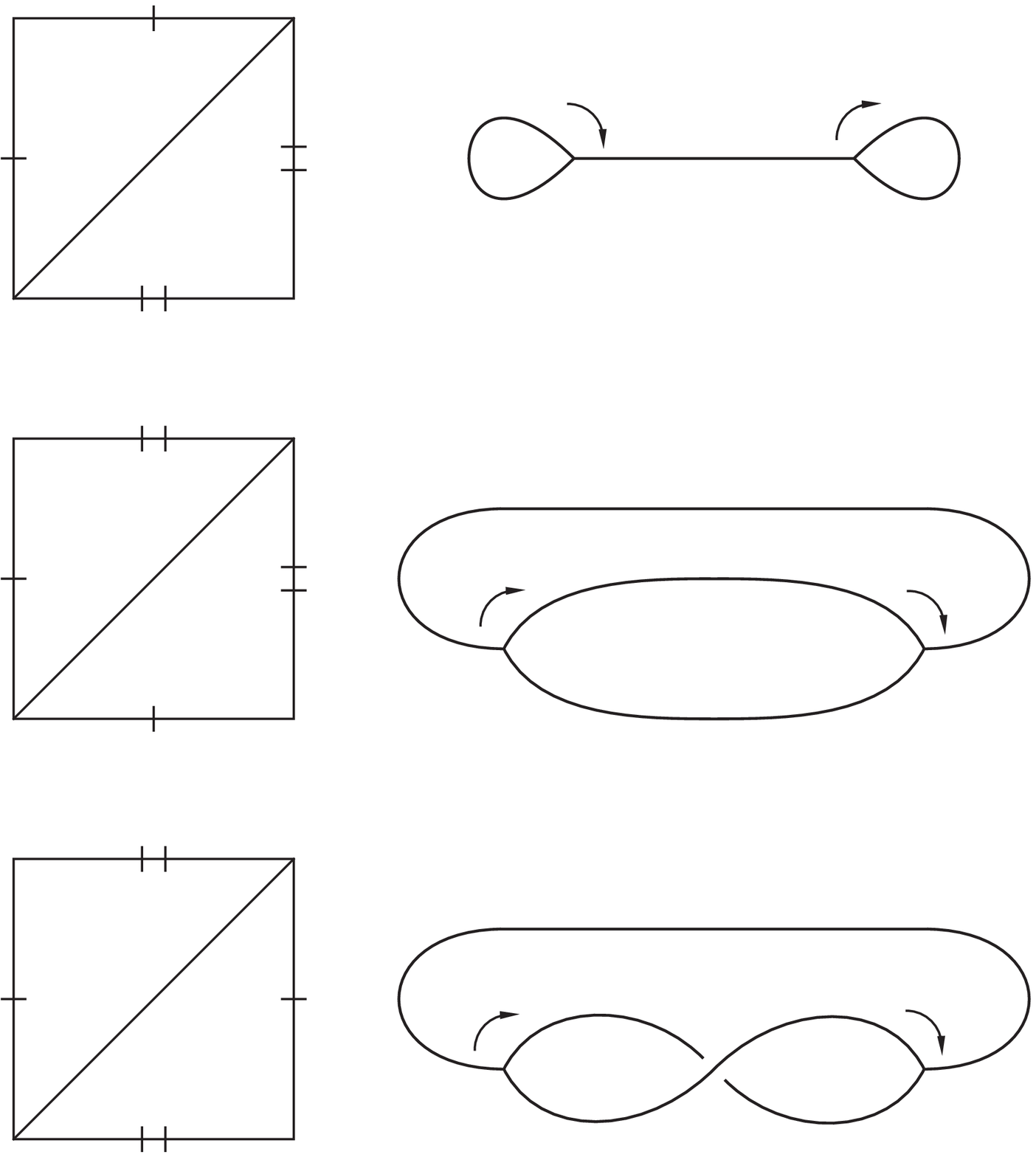}
\endinsert
\noindent
\centerline{Figure 1}
\vfill \eject 

\midinsert
\vskip .1in
\hskip 0in
\epsfxsize=5.2in
\epsfysize=6.8in
\epsfbox{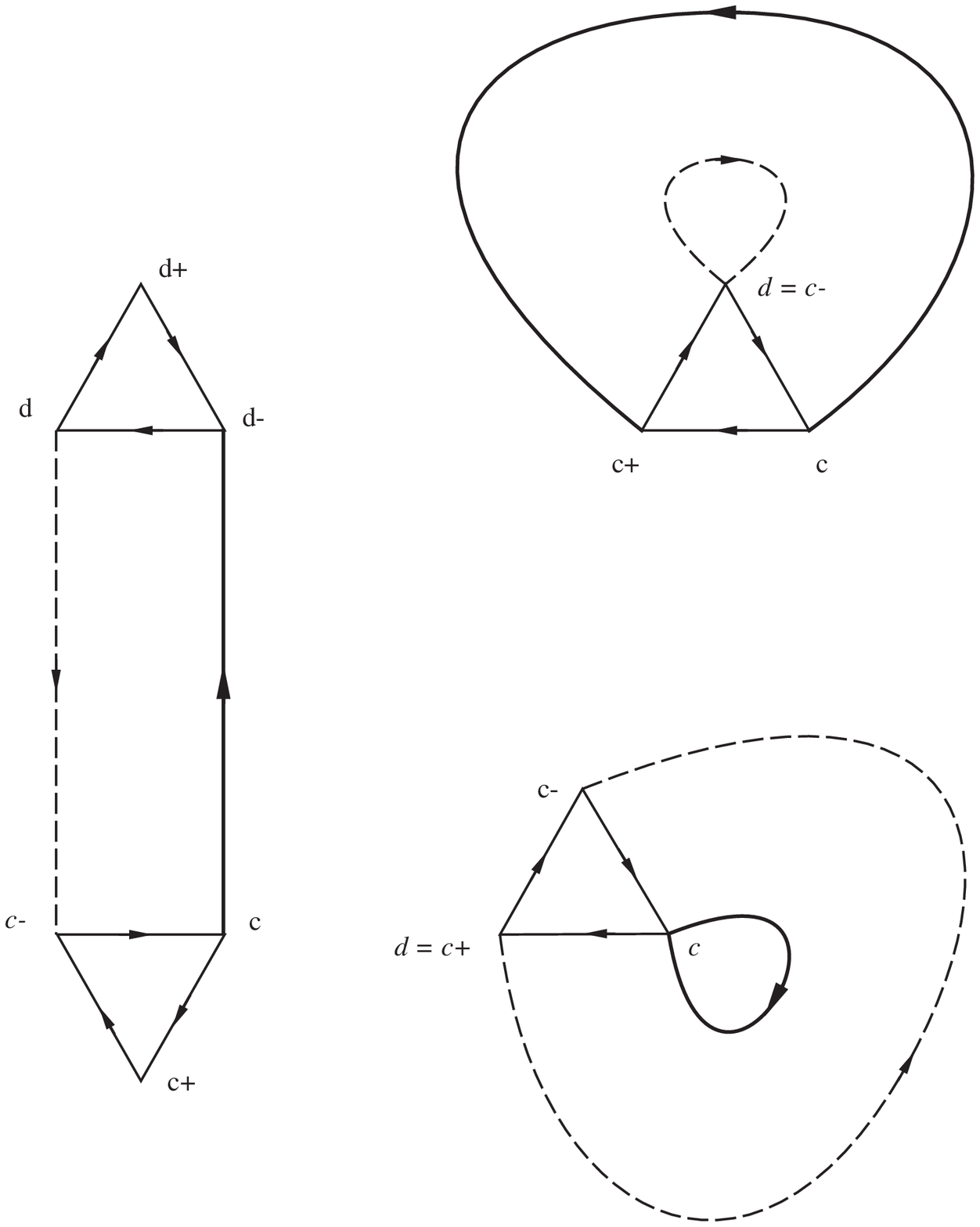}
\endinsert

\noindent
\centerline{Figure 2}
\vfill \eject 
\midinsert
\vskip .1in
\hskip 0in
\epsfxsize=5.2in
\epsfysize=6.8in
\epsfbox{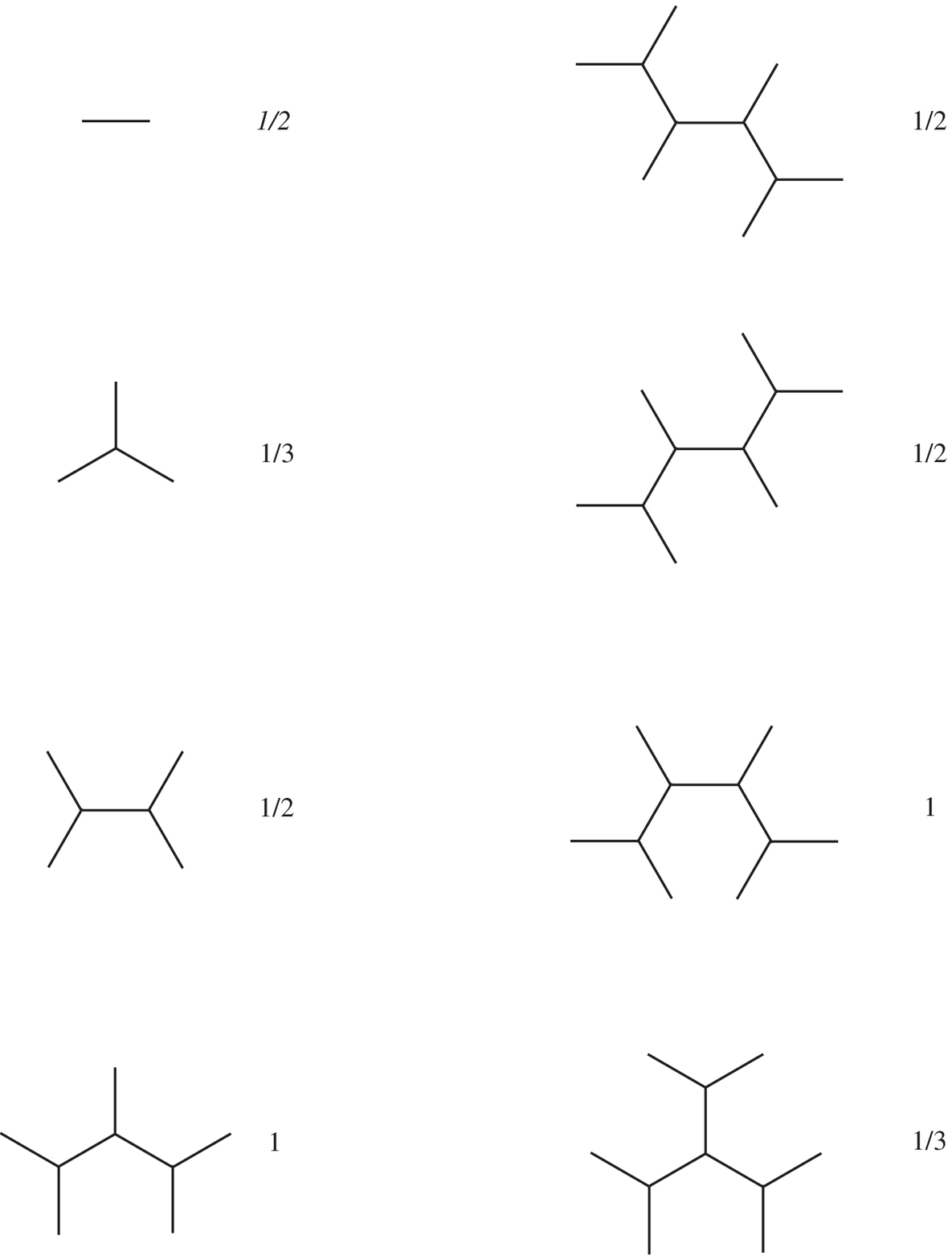}
\endinsert

\noindent
\centerline{Figure 3}
\vfill \eject 

\bye